\documentclass[a4paper,12pt]{article}
\usepackage{amsmath}
\usepackage{amsfonts}
\usepackage{bm}
\usepackage{amssymb}
\usepackage{graphicx}
\numberwithin{equation}{section}

\newcommand{\de}{\partial}

\newcommand{\rmd}{\textrm{d}}
\def\be{\begin{equation}}
\def\ee{\end{equation}}

\newcommand{\gam}{\gamma}

\newcommand{\del}{\delta}

\newcommand{\eps}{\varepsilon}

\renewcommand{\th}{\theta}

\newcommand{\om}{\omega}

\newcommand{\ens}[1]{\left \langle #1 \right \rangle}

\newcommand{\n}{{\hat{n}}}
\newcommand{\m}{\vec{m}}
\renewcommand{\k}{\vec{k}}
\renewcommand{\l}{\vec{l}}

\newcommand{\p}{\vec{p}}
\newcommand{\x}{\vec{x}}

\begin{document}
\begin{flushright} {\footnotesize ULB-TH/09-18}  \end{flushright}
\def\thefootnote{\fnsymbol{footnote}}

\begin{center}
\Large{\textbf{Sachs-Wolfe at second order: \\ the CMB bispectrum on large
    angular scales}}
\\[0.5cm]
\large{Lotfi Boubekeur$^{\rm a,b,c}$, Paolo Creminelli$^{\rm c}$, Guido D'Amico$^{\rm d,e}$,\\[.1cm]
Jorge Nore\~na$^{\rm d}$, and Filippo Vernizzi$^{\rm f}$}
\\[0.5cm]

\small{
\textit{$^{\rm a}$ Departament de F\'isica Te\`orica and IFIC,\\
Universitat de Val\`encia-CSIC, E-46100, Burjassot, Spain.}}

\vspace{.2cm}

\textit{$^{\rm b}$ Service de Physique Th\'eorique, Universit\'e Libre de Bruxelles, 1050 Brussels, Belgium}

\vspace{.2cm}

\small{
\textit{$^{\rm c}$ Abdus Salam International Centre for Theoretical Physics\\ Strada Costiera 11, 34014, Trieste, Italy}}

\vspace{.2cm}

\small{
\textit{$^{\rm d}$ SISSA, via Beirut 2-4, 34014, Trieste, Italy}}

\vspace{.2cm}

\small{
\textit{$^{\rm e}$ INFN - Sezione di Trieste, 34014 Trieste, Italy}}

\vspace{.2cm}

\small{
\textit{$^{\rm f}$ CEA, IPhT, 91191 Gif-sur-Yvette c\'edex, France\\ CNRS, URA-2306, 91191 Gif-sur-Yvette c\'edex, France}}

\end{center}

\vspace{.8cm}

\hrule \vspace{0.3cm}
\noindent \small{\textbf{Abstract}} \\[0.3cm]
\noindent
We calculate the 
Cosmic Microwave Background anisotropy 
bispectrum on large angular scales 
in the absence of primordial non-Gaussianities,
assuming exact matter dominance and extending
at second order the classic Sachs-Wolfe result $\delta T/T
=\Phi/3$. The calculation is done in Poisson gauge.
Besides intrinsic contributions calculated at last scattering, one must consider integrated effects. These are associated to
lensing, and to the time dependence of the potentials (Rees-Sciama) and of
the vector and tensor components of
the metric generated at second order. 
The bispectrum is explicitly computed in 
the
flat-sky
approximation. 
It
scales as $l^{-4}$ in
the scale invariant limit and 
the shape dependence of its various
contributions is represented in 3d plots.
Although all the contributions to the bispectrum are parametrically of the same order, the full bispectrum is dominated by lensing. In the squeezed limit it corresponds to
$f_{\rm NL}^{\rm local} = -1/6 - \cos(2 \theta)$, where $\theta$ is the
angle between the short and the long modes; the angle dependent
contribution comes from lensing. In the equilateral limit it
corresponds to $f_{\rm NL}^{\rm equil} \simeq 3.13 $.

\vspace{0.5cm} \hrule
\def\thefootnote{\arabic{footnote}}
\setcounter{footnote}{0}
\newpage

%%%%%%%%%%%%%%%%%%%%%%%%%%%%%%%%%%%%%%%%%%%%%%%%%%%%%%%%%%%%%%%%%%%%%%%%%%%%%%%%%%%
\section{Introduction}
%%%%%%%%%%%%%%%%%%%%%%%%%%%%%%%%%%%%%%%%%%%%%%%%%%%%%%%%%%%%%%%%%%%%%%%%%%%%%%%%%%%

The linear approximation to cosmological perturbations has been so far
sufficient and extremely fruitful, at least on large scales, before
non-linearities induced by gravity become significant. However, the
accuracy of observations is now reaching a level such that all second-order effects, naively of magnitude $\sim (10^{-5})^2$, may become
relevant. This is particularly important in the context of primordial
non-Gaussianities: second-order effects are in fact expected to give a
signal of order $f_{\rm NL} \sim$ few, which is not far from the
present experimental limits
\cite{Komatsu:2008hk,Slosar:2008hx,Smith:2009jr}. A large amount of
work has been done to study Cosmic Microwave Background (CMB)
fluctuations beyond the linear approximation, in order to make
predictions for the temperature bispectrum. As a complete calculation
of the bispectrum is a daunting task, people concentrated on specific
effects which are expected to dominate in particular limits. The
bispectrum generated by the correlation between lensing and the
Integrated Sachs-Wolfe (ISW) effect has been studied in \cite{Seljak:1998nu,Goldberg:1999xm}. The one coming from lensing and the Sunyaev-Zel'dovich effect has been studied in  \cite{Goldberg:1999xm}. In \cite{Khatri:2008kb,Senatore:2008vi,Senatore:2008wk} the bispectrum generated by perturbations in the recombination history has been calculated. Refs \cite{Pitrou:2008ak,Bartolo:2008sg} (see also \cite{Bartolo:2006fj}) focused on very short angular scales where the signal is dominated by the non-linearity induced by dark matter clustering. A systematic control of all second-order effects in the Boltzmann equations is currently under study: see \cite{Nitta:2009jp,Pitrou:2008hy} and references therein.

%%%%%%%%%%%%%%%%%%%%%%%%%%%%%%%%%%%
In this paper we calculate the CMB bispectrum in the limit of large
angles, i.e.~on angular scales larger than the one subtended by the
Hubble radius at recombination ($\theta \gtrsim 1^{\rm o}$); we do this assuming perfect matter
dominance. Important, although, as we will see, partial results were
obtained in this regime in \cite{Bartolo:2004ty,Bartolo:2005fp,Bartolo:2005kv}. 

Our calculation can be seen as the extension to second
order of the classic Sachs-Wolfe formula \cite{Sachs:1967er}
\be
\frac{\delta T}{T} = \frac{\Phi_e}{3} \label{SachsWolfe}\;,
\ee
where $\Phi_e$ is the Newtonian potential at recombination, which gives the large-angle prediction for the spectrum of the CMB fluctuations.
As it is well known, this formula describes the angular variation of the temperature
without considering the dynamics of the photon/baryon plasma, but only
the gravitational redshift of photons from the last scattering to
us. Therefore, it describes correctly the CMB anisotropies only in the
limit where the scales considered are well out of the Hubble radius at
recombination: the same restriction will apply to our calculation. The
Sachs-Wolfe formula further assumes that decoupling took place when
the universe was matter dominated -- neglecting the transition
between radiation and matter domination -- and that the universe is still
matter dominated nowadays, neglecting the present acceleration.
At linear order this simplification is very convenient
as the gravitational potential stays constant during matter dominance. At second-order the gravitational potential is no longer constant but the second-order metric during matter dominance is known \cite{Matarrese:1997ay} and can be written analytically as a function of the large-scale inflationary perturbations \cite{Bartolo:2005kv,Boubekeur:2008kn}. 

Clearly, these approximations do not hold in the real
universe. However, our calculations give the exact bispectrum in the
same limit in which the Sachs-Wolfe formula becomes exact: zero
cosmological constant, recombination that happens much after
equality and in the limit in which all scales are much larger than the
horizon at recombination. This last limit can be imagined by thinking
about an experimentalist making measurements in the far future, when
the angle subtended by the Hubble radius at recombination is minuscule.
The fact that our results become exact in a well defined physical
limit is quite important, as on large angular scales the separation
among different effects is in general gauge dependent. Therefore, one has to be
careful in making approximations because neglecting some effects leads,
in general, to a gauge dependent result.
Besides its theoretical interest, we expect our result to represent a fair approximation to the real
universe on large angular scales and it can be taken as a starting
point for more elaborated calculations.

Motivated by inflation, we assume that there are no vector or
tensor perturbations in the initial conditions on super-Hubble
scales. We perform the calculation of the CMB anisotropies by
integrating the photon geodesic equation during matter dominance using
the so called ``generalized Poisson gauge'', which generalizes at
second order the standard Newtonian gauge.
Besides the Newtonian and curvature potential, at second order new
terms are present in the metric, generated by the product of linear
fluctuations: a vector mode in the $\rmd x^i \rmd t$ entry of the metric, and
a tensor mode in the spatial part.

All these terms contribute to the final CMB anisotropy.
The time independent parts of the gravitational potentials give rise to second-order terms evaluated
at last scattering, in analogy with eq.~(\ref{SachsWolfe}); their contribution
was calculated in \cite{Bartolo:2004ty}. However, at second-order
there are also terms integrated along the photon trajectory, similarly
to what happens at first order when we depart from matter dominance
with the ISW effect. The time-evolution at second
order of the gravitational potential on sub-Hubble scales generates
the well-known Rees-Sciama effect \cite{Mollerach:1995sw,Munshi:1995eh}. But also the vector and tensor part
of the metric contribute with two integrated terms.\footnote{The
  integrated tensor contribution has been taken into account for the
  large scale anisotropies in \cite{Bartolo:2005kv}. The vector contribution has, to our knowledge, always been
ignored.} All these terms
contain a number of spatial gradients higher than the intrinsic terms,
so that one may think that they can be neglected on large scales  as
suppressed by positive powers of $k/(aH)$ at recombination. However,
this conclusion is too hasty:  these terms are integrated along the
photon trajectory while modes progressively reenter the Hubble radius. Thus
the ratio $k/(a H)$ should not be evaluated at recombination but when
the terms contribute to the time integral. We will see that all the
integrated pieces give a contribution of the same order as the
intrinsic terms in the equilateral limit. Actually the separation
between intrinsic and integrated effects has no physical, gauge
invariant meaning. For example, a part of the integrated vector
contribution will turn out to be a boundary term.

Another integrated contribution is gravitational lensing, due to the
gravitational deflection of the photon trajectory with respect to the
line of observation. Although the effect of lensing on the bispectrum
through its correlation with the ISW effect is well known
\cite{Seljak:1998nu,Goldberg:1999xm} (but absent in our calculation as
we are assuming perfect matter dominance), we will see that lensing is
important also when correlated with intrinsic contributions at last
scattering. In particular, we will find that lensing gives a squeezed
limit contribution of the same order as the one due to intrinsic
effects, but which depends on the angle between the long and the short
modes. The effect of lensing on the bispectrum was studied in
\cite{Creminelli:2004pv} with the conclusion that its effect is
suppressed in the squeezed limit by the tilt of the spectrum. We will
see that this conclusion is not correct.

In computing the CMB bispectrum we will employ the flat-sky
approximation, which is valid for small angles of view. Given that at
the same time we are interested in angles which are much larger than
the Hubble radius at recombination, there is a quite narrow range of
scales, $1 \ll l \ll l_{\rm 1st\; peak}$, where our approximations hold. However,
the flat-sky approximation greatly simplifies the algebra and makes
the result much more transparent. The results will be given by
2-dimensional kernels $B(\l_1, \l_2, \l_3)$, which can be thought of as
the 2d observable analogue of the kernels used (in 3 dimensions) to describe the
shape of the primordial non-Gaussianity \cite{Babich:2004gb}.

The paper is organized as follows. In the next section we give the
second-order metric in matter dominance in the generalized Poisson gauge as a function of the inflationary
initial conditions and we calculate
the temperature anisotropy at second order integrating the photon
geodesic. In section \ref{bispectrumshape} we make a general
discussion about the bispectrum of the temperature anisotropy in the
flat-sky approximation and we calculate this quantity induced by a
primordial non-Gaussianity of the local and equilateral kind. These
are useful for comparison with our results. In
section \ref{bispectrumcompute} we calculate the bispectrum using the
results of section \ref{SWsecord}. The calculation is split (for
convenience, not because the effects are physically distinguishable) in various pieces: intrinsic effects at last scattering,
integrated vector contribution, integrated tensor contribution and
lensing. The resulting total bispectrum is discussed in section \ref{bisfinal}
and conclusions are drawn in section \ref{conclusions}. The flat-sky
approximation is discussed in appendix \ref{app:flat_sky},
while the details of the calculation of the Rees-Sciama effect are
presented in appendix \ref{app:RS}.

%%%%%%%%%%%%%%%%%%%%%%%%%%%%%%%%%%%%%%%%%%%%%%%%%%%%%%%%%%%%%%%%%%%%%%%%%%%%%%%%%%%
\section{\label{SWsecord}Second-order temperature anisotropies}
%%%%%%%%%%%%%%%%%%%%%%%%%%%%%%%%%%%%%%%%%%%%%%%%%%%%%%%%%%%%%%%%%%%%%%%%%%%%%%%%%%%

In this section we calculate the
CMB temperature anisotropy  at second order in perturbations, 
in the large angular scale limit and for matter dominance, 
as a function of the angle of observation. 
On large angular scales, the effect of second-order perturbations on
the CMB fluctuations have been studied more generally in
\cite{Pyne:1995bs,Mollerach:1997up}. Although we will later use the
flat-sky approximation, the results of this section hold also in a
full-sky treatment.

We are interested in the CMB temperature fluctuations, 
\be
\label{deltaT/T}
\frac{\delta T }{T} (\n) \equiv \frac{T_o(\n) -\bar T_o}{\bar T_o} \;, 
\ee
where $T_o(\n)$ is the observed photon temperature in the angular
direction $\n$ ($\n^2 =1$) and $\bar T_o$ is its average over the
sky. For a black-body spectrum the observed temperature $T_o(\n)$ is
related to the one of emission $T_e(\x_e)$ by Liouville's theorem: as
phase space density is conserved in the propagation of photons
(assuming there is no further scattering), the phase space
density received in a given direction $\n$ is the same as the one at emission but with
a temperature \cite{Misner:1974qy,Sachs:1967er} 
\be
\label{T1}
T_o (\n) = \frac{\omega_o }{\omega_e} T_e (\x_e) \;,
\ee
where $\omega_e$ and $\omega_o$ are the frequencies at emission and
observation of a given photon. Notice that this statement is exact and
therefore holds at any order in perturbation theory. In general, also
the temperature at emission will not be isotropic, but will depend on
the angle of emission. This dependence can be however neglected in our
case, as we are interested in perturbations which are much longer than
the horizon at recombination.

We work in the so called generalized Poisson gauge and use conformal time $\tau$. In this gauge, the metric reads \cite{Matarrese:1997ay}
\begin{equation}
  \rmd s^2 = a^2(\tau) \left\{ - (1+2\Phi) \rmd \tau^2 
    + 2 \om_i \rmd x^i \rmd \tau
  + [(1-2\Psi) \del_{ij} + \gam_{ij}] \rmd x^i \rmd x^j \right\} \, ,
\label{metric}
\end{equation}
where $\omega_i$ is transverse, $\om_{i, i} =0$, and $\gam_{ij}$ is
transverse and traceless, $\gam_{ij,i} = 0 = \gam_{ii}$.
In the matter dominated era, assuming that the amount of primordial
gravitational waves is negligible, the components of this metric are \cite{Matarrese:1997ay,Bartolo:2005kv,Boubekeur:2008kn}
\begin{align}
  \Phi = & \phi+
  \left[ \phi^2 + \de^{-2}(\de_j\phi)^2
    -3\de^{-4}\de_i\de_j(\de_i\phi\de_j\phi) \right] \nonumber \\
  & +\frac{2}{21a^2H^2}\de^{-2}\left[2(\de_i\de_j\phi)^2
    +5(\de^2\phi)^2+7\de_i\phi\de_i\de^2\phi\right] \, , \label{Phi} \\
  \Psi = & \phi -\left[\phi^2 + \frac{2}{3} \de^{-2}(\de_i\phi)^2
    - 2 \de^{-4} \de_i\de_j(\de_i\phi\de_j\phi) \right] \nonumber \\
  & + \frac{2}{21 a^2 H^2} \de^{-2}\left[2(\de_i\de_j\phi)^2
    +5(\de^2\phi)^2+7\de_i\phi\de_i\de^2\phi\right] \, , \label{Psi} \\
  \om_{i} =& - \frac{8}{3 a H} \de^{-2} \left[\de^2 \phi \de_i \phi
    - \de^{-2} \de_i\de_j 
    (\de^2\phi \de_j\phi) \right] \, , \label{omega} \\
  {\gamma}_{ij} = & - 20 
  \left( \frac{1}{3} - \frac{j_1(k\tau)}{k \tau} \right)
  \de^{-2} P_{ij \, kl}^{\rm TT}
  \left( \de_k\phi\de_l\phi \right) \label{gamma} \, .
\end{align}
The scalar quantities $\Phi$ and $\Psi$ are the Newtonian and
curvature potentials, respectively, while we will refer to $\omega_i$
and $\gamma_{ij}$ as the vector and tensor components of the
metric. The metric is expressed in terms of
$\phi$, the time-independent quantity representing the initial
curvature perturbation generated during inflation. Indeed, $\phi$ is simply
proportional to the (non-linear) curvature perturbation on uniform density hypersurfaces $\zeta$: on super-Hubble scales, where $\zeta$ is constant, 
\be
\label{zeta_phi}
\phi = -\frac{3}{5} \zeta  \qquad (k \ll aH) \;. 
\ee
In the following we are going to assume that $\zeta$ on large scales, and therefore
$\phi$, is perfectly Gaussian, which is a very good
approximation for example in minimal single field inflationary models \cite{Maldacena:2002vr,Acquaviva:2002ud}. 
In the expression for tensor modes, the spherical Bessel function
$j_1(x)$ is given by $j_1(x)=\sin(x)/x^2-\cos(x)/x$, while $P_{ij \, kl}^{\rm TT}$ is a transverse traceless projector defined as
\begin{equation}
\label{projector}
P_{ij \, kl}^{\rm TT}\equiv \frac12 \left(P_{ik} P_{jl} + P_{jk} P_{il}- P_{ij} P_{kl}\right) \;,
\end{equation}
where $P_{ij}$ is a symmetric transverse projector given by
\begin{equation}
P_{ij}\equiv \delta_{ij} -{\partial_i\partial_j\over \partial^2} \;.
\end{equation}
It can be expanded to give 
\begin{equation}\label{expansion2}
  P_{ij \, kl}^{\rm TT} \left( \de_k\phi\de_l\phi \right) = - \de^{-2}
  \left[ \de^2 \Theta_0 \delta_{ij} + \de_i \de_j \Theta_0 +2 (\de^2 \phi
    \de_i \de_j \phi - \de_i \de_k \phi \de_j \de_k \phi) \right]\;,
\end{equation}
with
\begin{equation}\label{Theta}
  \Theta_0 = - \frac12 \de^{-2} \left[ (\de^2 \phi)^2 -(\de_i\de_j\phi)^2 
  \right] \, .
\end{equation}

In order to study the photon redshift we must solve the photon geodesic equation from last scattering to us, taking into account the perturbations of the metric above. The photon geodesic equation can be written as
\begin{equation}
  \frac{\rmd P_\mu}{\rmd\lambda} = \frac{1}{2}\partial_\mu g_{\alpha\beta}P^\alpha
  P^\beta \, , \label{geod}
\end{equation}
where $P^\mu = dx^\mu /d \lambda$ is the four-momentum of the photon, $P^\mu P_\mu =0$. 
The frequency of a photon with four-momentum $P^\mu$ as measured by an observer with four-velocity $u^\mu$, is given by $\omega = -P_\mu u^\mu$. 
For simplicity, we choose the observer today to have zero spatial velocity, $u_o^i=0$. Indeed, any peculiar motion of the observer leads to a dipole anisotropy that can easily be subtracted. Furthermore, since we are interested in the large angular scales, we neglect also the Doppler 
effect due to the velocity of the photon/baryon fluid at
recombination, which vanishes on super-Hubble scales at
decoupling. Thus, we choose also the emitter to have zero spatial
velocity, $u_e^i =0$, so that we have $\omega = - P_0 u^0 $ both for the observer and the emitter.
Making use of the normalization condition of the four-velocity, $u^\mu u_\mu =-1$, one obtains 
$\omega = -P_0 / \sqrt{-g_{00}}$, and thus
\be
\label{omega_P0}
\frac{\omega_o}{\omega_e}=  \frac{P_0 (\tau_o)}{P_0 (\tau_e)} \frac{ \sqrt{-g_{00}}|_e }{  \sqrt{-g_{00}}|_o} \;.
\ee

In order to compute $P_0$ we need to solve the
time component of eq.~\eqref{geod}. Using that $P^0 = \rmd\tau/\rmd\lambda$ and plugging the metric
\eqref{metric} into eq.~\eqref{geod} yields
\begin{equation}
P^0  \frac{\rmd P_0}{\rmd \tau} = \mathcal{H}g_{\alpha\beta}P^\alpha
  P^\beta - a^2\Phi' (P^{0})^2 + a^2\omega_i'P^0 P^i +
  a^2\Big(-\Psi'\delta_{ij} + \frac{1}{2}\gamma'_{ij}\Big)P^i P^j\,,
\end{equation}
where by a prime we denote the partial derivative with respect to conformal time, ${}' \equiv  \partial / \partial \tau$, and $\cal H$ is the conformal Hubble rate, ${\cal H} \equiv a' /a$.
One can immediately notice that the first term on the right hand side
vanishes because of the massless condition $P^\mu P_\mu=0$. Note also
that, as we are studying perfect matter dominance, the two potentials
$\Phi$ and $\Psi$ are constant at linear order, see eqs.~(\ref{Phi}) and (\ref{Psi}). Thus, their time derivatives $\Phi'$, $\Psi'$, together with $\omega_i'$ and
$\gamma_{ij}'$, are all second-order quantities. One can
therefore replace the zeroth-order expression $P^i = - P^0 \hat n^i$ into this equation; furthermore, using the background relation $P^0 = -P_0/a^2$,
the geodesic equation can be finally rewritten as
\begin{equation}
\frac{1}{P_0}  \frac{\rmd P_0}{\rmd \tau} = \Phi' + \Psi' + \omega_i'\hat{n}^i -
  \frac{1}{2}\gamma_{ij}'\hat{n}^i\hat{n}^j ,
\end{equation}
that upon integration yields
\begin{equation}
\frac{P_0(\tau_o)}{P_0(\tau_e)}  =  1+ \int_{\tau_e}^{\tau_o}\rmd\tau\;\Big(\Phi' +
  \Psi' + \omega_i'\hat{n}^i -
  \frac{1}{2}\gamma_{ij}'\hat{n}^i\hat{n}^j\Big)\,.
\label{geodesic}
\end{equation}
Plugging this expression into eq.~(\ref{omega_P0}), one obtains the photon redshift up to second-order as a function of the metric perturbations,
\begin{equation}
 \frac{ \omega_o}{ \omega_e}=  \frac{a_e}{a_o} \sqrt{\frac{1+2 \Phi_e}{1+2 \Phi_o}}
 \Bigg[1  +
  \int_{\tau_e}^{\tau_o}\rmd\tau\, \Big(\Phi' + \Psi' +
  \omega_i'\hat{n}^i -
  \frac{1}{2}\gamma_{ij}'\hat{n}^i\hat{n}^j\Big)\Bigg] \;.
  \label{freq}
\end{equation}

Now we need to relate $T_e( \x_e)$ on the right hand side of eq.~(\ref{T1}) to the metric perturbations at decoupling. Since we concentrate on large angular scales, we only need the super-Hubble relation. We will use adiabatic initial conditions. In this case the dark matter energy density $\rho_m$ simply scales as the third power of the temperature,
\be
\rho_m \propto T_e^3 \;. 
\label{adiabatic}
\ee
In the matter dominated era, the energy density of dark matter is related to the metric perturbations through the Einstein equations, in particular through the energy constraint equation. On super-Hubble scales, i.e.~neglecting spatial gradients, and using the fact that the potentials $\Phi$ and $\Psi$  at first order are time-independent in the matter dominated era this reads, up to second order, (see for instance eq.~(196) of \cite{Langlois:2006vv})
\be
3 H^2 \left( 1 -2 \Phi_e + 4 \Phi_e^2 \right) = 8 \pi G \rho_m \;,
\ee
where $H$ is the Hubble rate. Using the background Friedmann equation and eq.~(\ref{adiabatic}) above, this equation can be rewritten as
\be
\label{TeEinstein}
T_e = \left( 1 -2 \Phi_e + 4 \Phi_e^2 \right)^{1/3} \bar T_e \;,
\ee
where $\bar T_e$ is the average temperature at emission, which simply
scales as the inverse of the background scale factor. 

%%%%%%%%%%%%%%%%%%%%%%%%%%%%%%%%%%%%%%%%%%%%%%%%%%%%%%%%%%%%%%%%%%%%%%%%%%%%%%%%%%%

This equation can be derived in a simpler way \cite{Bartolo:2005fp}
taking into account that, at recombination, all the modes that we are
considering are much longer than the horizon and adiabatic. This means
that each local observer will see a completely unperturbed history at
any order in perturbations. Indeed, the vector and tensor components of
the metric, eqs.~(\ref{omega}) and (\ref{gamma}), are suppressed
by powers of $k/(aH)$ and can be neglected at recombination.\footnote{Notice that in eq.~(\ref{gamma}) the
  prefactor in parentheses, $1/3- j_1(k \tau)/(k \tau)$, goes to zero for $k \tau \to 0$, i.e.~when the $\gamma$ mode is out of the
  horizon.} The same holds for the time dependent part of $\Phi$ and
$\Psi$, i.e.~the second lines of eq.~(\ref{Phi}) and
(\ref{Psi}). This means that the metric on large scales takes the form
\begin{equation}
  \rmd s^2 = a^2(\tau) \left\{ - (1+2\Phi) \rmd \tau^2    
  + (1-2\Psi) \del_{ij} \rmd x^i \rmd x^j \right\} \, , \qquad (k\ll aH) \;,
\end{equation}
where $\Phi$ and $\Psi$ are now time independent and slowly varying in
space. Locally, i.e.~on scales of order of the horizon at recombination, this metric describes an unperturbed universe as the
terms with $\Phi$ and $\Psi$ can be taken to be constant in space and reabsorbed with a change of
coordinates. In particular, the evolution is unperturbed in terms of a
new conformal time $\tilde \tau$ which satisfies (in matter dominance
$a \propto \tau^2$)
\be
\tau^4 (1+2 \Phi) d \tau^2 = \tilde\tau^4 d\tilde\tau^2\;,
\ee
i.e.~$\tilde\tau = \tau (1+2\Phi)^{1/6}$. The temperature on a $\tau
= {\rm const.}$ surface will be perturbed, because the same value of
$\tau$ corresponds to different moments along the unperturbed
evolution, i.e.~to different values of $\tilde\tau$. As $T  \propto
1/\tilde \tau^2$ we have
\be
T_e = (1+ 2 \Phi_e)^{-1/3} \bar T_e \;,
\ee  
which coincides, at second order, with eq.~(\ref{TeEinstein}).
 
%%%%%%%%%%%%%%%%%%%%%%%%%%%%%%%%%%%%%%%%%%%%%%%%%%%%%%%%%%%%%%%%%%%%%%%%%%%%%%%%%%%%%

Now, let us plug both this equation and eq.~(\ref{freq}) into eq.~(\ref{T1}) and write the observed CMB temperature up to second-order as a function of the metric perturbations, 
\be
T_o (\n) =  \frac{a_e}{a_o} \bar{ T}_e \sqrt{\frac{1+2 \Phi_e}{1+2 \Phi_o}}
 \left( 1 + 2 \Phi_e \right)^{-1/3} \Bigg[1  +
  \int_{\tau_e}^{\tau_o}\rmd\tau\, \Big(\Phi' + \Psi' +
  \omega_i'\hat{n}^i -
  \frac{1}{2}\gamma_{ij}'\hat{n}^i\hat{n}^j\Big)\Bigg] \;.
\ee
Note that on the right hand side of this equation, the gravitational potential at the observer, $\Phi_o$, does not depend on the direction of observation. Thus, its dependence can be simply reabsorbed into the definition of $\bar T_o$. Expanding this equation up to second order in the perturbation and plugging the right hand side in eq.~(\ref{deltaT/T}) we finally obtain the CMB temperature anisotropies,
\be
\label{CMBani1}
\frac{\delta T_o }{T_o} (\n) =   \frac{1}{3} \Phi_e  - \frac{5}{18} \Phi_e^2 +
  \int_{\tau_e}^{\tau_o}\rmd\tau\, \Big(\Phi' + \Psi' +
  \omega_i'\hat{n}^i -
  \frac{1}{2}\gamma_{ij}'\hat{n}^i\hat{n}^j\Big)\;.
\ee

The first two terms on the right hand side of this equation have to be evaluated at the position of the emitted photon, $ \x_e$. Since the second term is second-order, it can be simply evaluated at the background position $\n D_e$, with $D_e \equiv \tau_o - \tau_e$. Also the integral is second-order; thus
it can be computed along the background photon trajectory, i.e.~$\x (\tau) = \n \, D(\tau)$,  $D(\tau) \equiv \tau_o - \tau$. However, the first term on the right hand side is a first-order quantity. Thus, at second order it must be evaluated at the perturbed position of the photon at emission. Expanding around the background position $\n D_e$ we can write it as 
\be
\label{expansion}
\Phi (\x_e)=  \Phi (\n D_e) + {\delta \x}_e \cdot  \vec{\nabla} \phi(\n D_e) \;,
\ee
where $\delta \x_e \equiv \x_e - \n D_e$ is the deviation from the background trajectory and we have used that $\Phi = \phi$ at first order. 

In order to find $\delta \x_e$ we must solve the spatial component of the geodesic equation. Since $\vec \nabla \phi$ is already first-order we need to compute $\delta \x_e$ at first-order only. Thus, equation (\ref{geod}) gives
\begin{equation}
P^0  \frac{\rmd P_i}{\rmd \tau} =  - 2 a^2 \partial_i \phi  (P^0)^2 \,,
\end{equation}
where we have used that $\Phi +\Psi =2 \phi$ at first order.
This equation can be integrated using the background relation $P^0 \propto 1/a^2$. The spatial gradient can be decomposed along and orthogonally to the background photon trajectory. Since $\phi$ is time-independent, the component along the photon trajectory is a total derivative. Furthermore, raising the spatial index with the first order metric and then using $P^0 \propto (1-2 \phi)/a^2$ one obtains
\be
  \frac{dx^i}{d \tau} = \frac{P^i}{P^0} = - \hat n^i (1+ 2 \phi) 
  + 2 \int_{\tau}^{\tau_o}\rmd\tau'  \nabla_\parallel^i \phi \;,
\ee
where we have defined $\nabla_\parallel^i \equiv (\delta^{ij} - \hat n^i \hat n^j) \partial_j $ as the spatial gradient orthogonal to the line of sight\footnote{Notice that the direction
perpendicular to the photon trajectory is parallel to the flat sky, 
so that, in our notation, the gradient is \emph{parallel to the sky}.}  and we have absorbed the dependence on
$\phi_o$ in the first-order definition of $\n$, $\hat n^i \equiv - P^i_o/P^0_o (1+2 \phi_o)$.
Integrating this equation and subtracting the background value $\n D_e$, after an integration by parts in the second integral one obtains the geodesic deviation
\be
\label{geodesic_deviation}
\delta \x_e = 2 \n \int_{\tau_e}^{\tau_o}\rmd\tau \phi - 2
\int_{\tau_e}^{\tau_o}\rmd\tau (\tau - \tau_e) \vec{\nabla}_\parallel
\phi\;.  \ee The first term on the right hand side, longitudinal to
the line of sight, is the so-called Shapiro time-delay. This effect
was discussed in \cite{Hu:2001yq} and we will discard it from the
following discussion where we will concentrate on modes much shorter
than the present Hubble radius, where the flat-sky approximation is
valid. Indeed, since the integral of $\phi$ tends to average to zero
unless the mode wave-vector is orthogonal to the line of sight, it
gives a negligible contribution to the CMB anisotropy for $l \gg
1$. The second term is the transverse deviation from the background
trajectory, responsible for the lensing effect \cite{Seljak:1995ve}.

Including the lensing effect by re-expressing $\Phi_e$ using eq.~(\ref{expansion}) and re-writing $\Phi$ in terms of $\phi$ using the large-scale limit of eq.~(\ref{Phi}), i.e.,
 $ \Phi = \phi + \phi^2 + \de^{-2}(\de_j\phi)^2 -
  3\de^{-4}\de_i\de_j(\de_i\phi\de_j\phi)$, eq.~(\ref{CMBani1}) can be
  finally written as
\begin{align}
  \frac{\delta T}{{T}}(\n) =   \,  & \left[ \frac{1}{3}\phi +   \frac{1}{18}\phi^2
  +\frac{1}{3}\de^{-2}\big((\de_i\phi)^2 -
  3\de^{-2}\de_i\de_j(\de_i\phi\de_j\phi)\big) \right]_e \nonumber \\
  +  & 
  \int_{\tau_e}^{\tau_o}\rmd\tau\, \Big(\Phi' + \Psi' +
  \omega_i'\hat{n}^i - \frac{1}{2}\gamma_{ij}'\hat{n}^i\hat{n}^j\Big)
  + \frac{1}{3} \vec \alpha  \cdot \vec \nabla_{\n} \phi_e  \,, 
  \label{CMBfinal}
\end{align}
where $\vec \alpha$ is the deviation angle given by eq.~(\ref{geodesic_deviation}) as
\be
\label{alpha2}
\vec \alpha \equiv - 2 \int_{\tau_e}^{\tau_o}\rmd\tau \frac{\tau - \tau_e}{\tau_o - \tau_e} \vec{\nabla}_\parallel \phi\;.
\ee
On the right hand side of eq.~(\ref{CMBfinal}), the subscript ``$e$'' means at the background position of the emitted photon, $\n D_e$.
The first line of eq.~(\ref{CMBfinal}) was found in \cite{Bartolo:2004ty}. It represents an intrinsic effect due to the combination of the Doppler effect and the adiabatic temperature fluctuation of the plasma at recombination. The second line contains the Rees-Sciama effect, due to the second-order time evolution of the scalar potentials, and the effect of the time dependence of the vector and tensor components of the metric. Finally, the last term in the second line of eq.~(\ref{CMBfinal}) represents the lensing effect. All these effects were discussed for a more general metric in \cite{Pyne:1995bs,Mollerach:1997up}. 
%

%%%%%%%%%%%%%%%%%%%%%%%%%%%%%%%%%%%%%%%%%%%%%%%%%%%%%%%%%%%%%%%%%%%%%%%%%%%%%%%%%%%%%%%

There is a nice way to check the factor $\phi_e^2/18$ in the expression
(\ref{CMBfinal}) which, as we will see, is important for the squeezed limit of the
bispectrum \cite{Bartolo:2004ty}. Let us take the limit in which one of the two Fourier modes of the initial
conditions $\phi_e$ becomes infinitely
long. This mode is still out of the horizon today and therefore cannot
affect any physical observable. Let us check that this is indeed the
case. When one of the wavevectors goes to zero, all the terms
containing spatial derivatives in the expression above vanish, as it is
clear from the explicit form of the metric eqs.~(\ref{Phi})--(\ref{gamma}). One is
left only with the first two terms which, up to second order, it is
useful to rewrite in an exponential form \cite{Bartolo:2005fp} as
\be
\label{exp}
  \frac{\delta T}{{T}}(\n) =   \,  \left[ \frac{1}{3}\phi +   \frac{1}{18}\phi^2
  \right]_e \simeq e^{\phi_e/3} -1 \;.
\ee
At first sight it looks as if the constant mode could affect observations
through the second order term, which mixes a short mode with the
constant one. This actually is not the case as the
constant mode also affects the average measured temperature. Indeed the
well defined measurable quantity is given by
\be
\label{expback}
\frac{T_o(\n) -\bar T_o}{\bar T_o} = \frac{e^{\phi_e/3}}{ \langle
e^{\phi_e/3}\rangle} -1 \;.
\ee
Now we see that indeed a constant contribution to $\phi_e$ cancels out:
the quadratic term cancels with the redefinition of the average
temperature. Notice that this is only possible because of the
exact numerical coefficient $1/18$ in front of the quadratic term. For
the calculation of the bispectrum we are only interested in modes
inside the Hubble radius at present time, thus it is not necessary to modify eq.~(\ref{CMBfinal}) to take into account the correct average temperature as in eq.~(\ref{expback}).

In this way we also understand why the argument presented in
\cite{Creminelli:2004pv} for the squeezed limit of the 3-point function is not correct. In that
reference it is argued that a term like $\phi_e^2/18$, which induces a
correlation between short and long modes, cannot exist,
as it would imply -- as in eq.~(\ref{exp}) -- that a mode which is
still out of the horizon gives a measurable effect. What was neglected
is that the same mode would change the average of the measured
temperature.

%%%%%%%%%%%%%%%%%%%%%%%%%%%%%%%%%%%%%%%%%%%%%%%%%%%%%%%%%%%%%%%%%%%%%%%%%%%%%%%%%%%%%%
\section{\label{bispectrumshape}The CMB bispectrum and its shape}
%%%%%%%%%%%%%%%%%%%%%%%%%%%%%%%%%%%%%%%%%%%%%%%%%%%%%%%%%%%%%%%%%%%%%%%%%%%%%%%%%%%%%%

In this section we will discuss the CMB bispectrum and its shape
dependence. We will use the flat-sky approximation. Even though this
approximation is not very good for the lowest multipoles, the
expressions that we will derive are much more transparent than using a
full-sky treatment.

In the flat-sky approximation (see appendix~\ref{app:flat_sky}) the Fourier transform in the sky of the temperature anisotropies is
\be
\label{a_l_ini}
a_{\l} = \int \rmd^2 m \,  \frac{\delta T}{T} (\n) \, e^{-i \l \cdot \m } \;,
\ee
where we have decomposed $\n$ into a part orthogonal 
and parallel to the line of sight as $\n \simeq (\m, 1) $ (see appendix~\ref{app:flat_sky}).
The spectrum of the 2-point function is defined as
\be
\label{C_l}
\langle a_{ \l} \, a_{\l'} \rangle = (2\pi)^2 \delta(\l + \l') C_{l}\;.
\ee
We can rewrite the standard linear Sachs-Wolfe term in eq.~(\ref{CMBfinal}) in Fourier space,
\be
\label{T_L}
\frac{\delta T}{T} (\n) 
=   \int \frac{\rmd^3 k}{(2 \pi)^3} \, \frac13 \phi_{\k}  \, e^{i \k \cdot \n D_e} \;.
\ee
As explained more accurately in appendix \ref{app:flat_sky}, it is
convenient to separate $\vec k$ as the sum of a 2-dimensional vector
parallel to the flat sky and a component orthogonal to it, 
\be
\vec k \equiv (\vec k^\parallel, k^\perp) \;. 
\ee
Using this decomposition and 
inserting eq.~(\ref{T_L}) in eq.~(\ref{a_l_ini}) one obtains
\be
\label{a_l}
a_{\l} =   \int \frac{\rmd^3 k}{(2 \pi)^3} \, \frac13 \phi_{\k}  \, e^{i k^{\perp} D_e} \, (2 \pi)^2\delta(\l - \k^\parallel D_e) \;. 
\ee
From this expression the power spectrum defined in eq.~(\ref{C_l}) reads,
\be
\label{C_lA}
C_{l} = \frac{A}{9 \pi l^2}\;,
\ee
where for simplicity we have used a scale invariant power spectrum for the gravitational potential $\phi$,
\be
\label{power_spectrum}
\langle \phi_{\k} \, \phi_{\k'} \rangle \equiv (2\pi)^3 \delta (\k + \k') \frac{A}{k^3} \;.
\ee

We are interested in the ensemble average of the product of three $a_{\l}$. Thus, we define the CMB bispectrum $B(\l_1, \l_2, \l_3)$ as
\be
\label{def_bispectrum}
\langle a_{\l_1} a_{\l_2} a_{\l_3} \rangle = (2\pi)^2 \delta(\l_1 + \l_2 + \l_3) B(\l_1, \l_2, \l_3) \;.
\ee
Translational and rotational invariance reduce the number of degrees
of freedom of $B$ to three independent variables only, for instance
$l_1, l_2, l_3 $. This is completely general, but in the particular
limit that we are studying (large scales and perfect matter dominance) we
will also see that the leading contributions to the bispectrum are scale invariant, i.e.~the amount of non-Gaussianity is the same at long and short
scales. Mathematically this implies that the function $B$ is a homogeneous function of degree $-4$, 
\be
\label{scaling}
B(\lambda \l_1,\lambda \l_2,\lambda \l_3) = \lambda^{-4} B(\l_1, \l_2, \l_3)\;,
\ee
which further reduces the number of degrees of freedom to two, for
instance the ratios $r_2 \equiv l_2/l_1$ and $r_3 \equiv
l_3/l_1$. Without loss of generality we can assume $0 \leq r_2 \leq
r_3 \leq 1$; the triangle inequality implies $r_2 \ge 1 - r_3$. 
This is very similar to what happens when one studies the shape
dependence of the primordial 3-point function of the curvature
perturbation \cite{Babich:2004gb}, with the difference that here we
are in two and not three dimensions.

We are interested in the dependence of $B$ on the two ratios $r_2$ and
$r_3$, which describes how the bispectrum changes as we change the shape
of the triangle in Fourier space. The possibility to measure a
bispectrum depends on its signal to noise ratio $S/N$, which is
given in flat-sky approximation by \cite{Hu:2000ee}
\be
\label{SN2}
(S/N)^2 = \frac{1}{\pi} \int \frac{\rmd^2 l_2 \rmd^2 l_3}{(2\pi)^2} \frac{B(\l_1,\l_2,\l_3)^2}{6 C_{l_1}C_{l_2}C_{l_3}}\;.
\ee
The overall scaling in $l$ is fixed by eq.~(\ref{scaling}) and (\ref{C_lA}): the integrand scales as
$l^{-2}$. To study the shape dependence one can look at the quantity
\be
\label{toplot}
r_2\,r_3\,B(1,r_2,r_3)\;.
\ee
The square of this quantity is in fact proportional to the integrand in the
expression above and thus quantifies the contribution to $(S/N)^2$ of triangles
with a given shape. To be more precise one could rewrite the
  expression (\ref{SN2}) for $(S/N)^2$ as an integral over the
  two ratios $r_2$ and $r_3$
\be
(S/N)^2 \propto \int \rmd r_2 \rmd r_3 \left[\frac{r_2^{3/2} r_3^{3/2}}{(2 r_2^2 +2
  r_3^2 +2 r_2^2 r_3^2 -1 -r_2^4 -r_3^4)^{1/4}} B(1,r_2,r_3)\right]^2 \;.
\ee
Therefore it would seem appropriate to consider the function in brackets
as a measure of the $S/N$ contribution; in this way in fact the
integral of the square of the function over an $r_2, r_3$ region would
directly give the contribution of those shape configurations to
$(S/N)^2$. This would exactly parallel what is done in \cite{Babich:2004gb} to study
the shape dependence of the primordial 3-point function. However in
this way we would introduce a spurious divergence in the
plots for flattened configurations when all the sides of the triangle
are aligned: indeed, the denominator of the expression above blows up
in this limit. This is just a consequence of describing the triangle
shape in terms of $r_2$ and $r_3$ and it does not imply that flattened
triangles are indeed more important. For this reason we prefer to plot
$r_2\,r_3\,B(1,r_2,r_3)$ in the following.

For comparison with the results that we will derive later, it is
interesting to study the function (\ref{toplot}) when the CMB
bispectrum is dominated by a primordial contribution. Two interesting
cases are given by the so-called local and equilateral shapes \cite{Babich:2004gb}.

\subsection{The local shape} 

\begin{center}
\begin{figure}[htc]
\begin{tabular}{cc}
\resizebox*{0.48\textwidth}{!}{\includegraphics{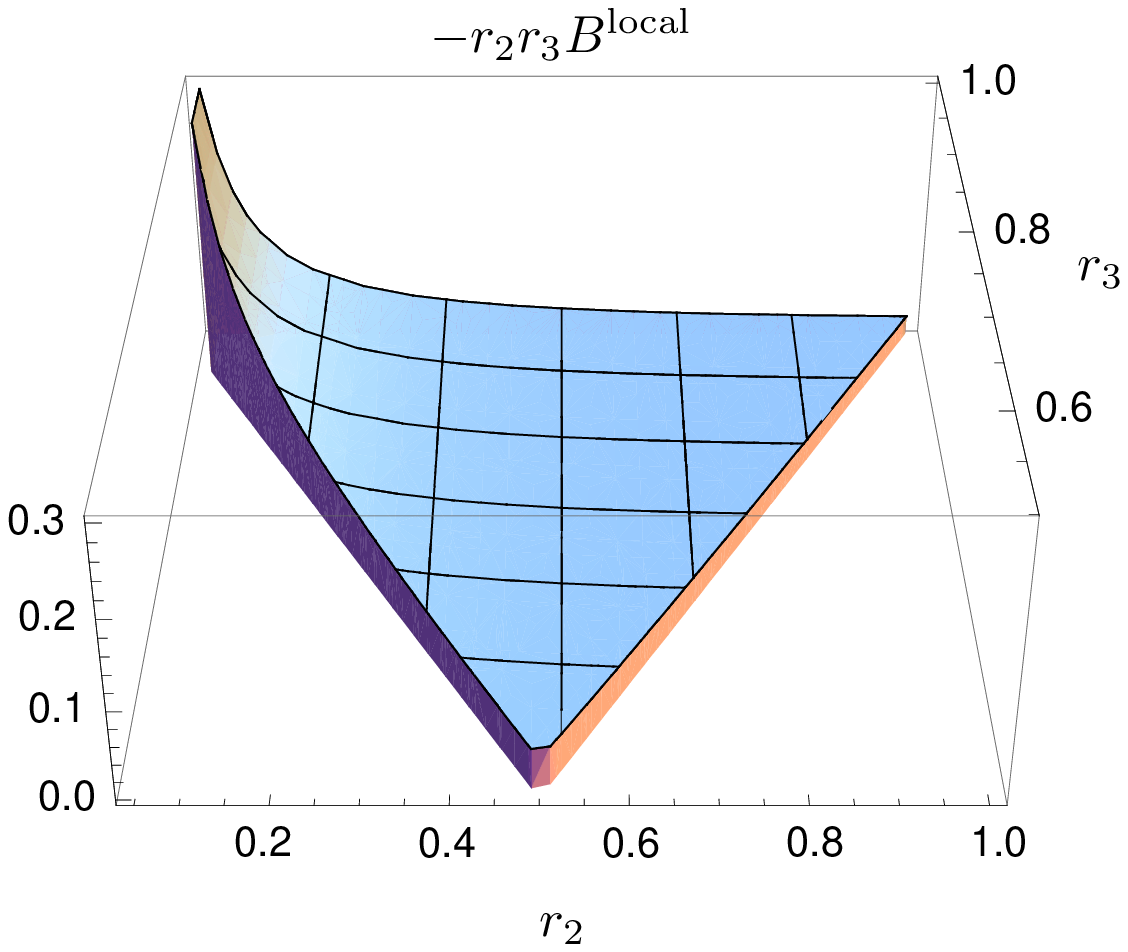}}   &
\raisebox{0.9cm}{\resizebox*{0.48\textwidth}{!}{\includegraphics{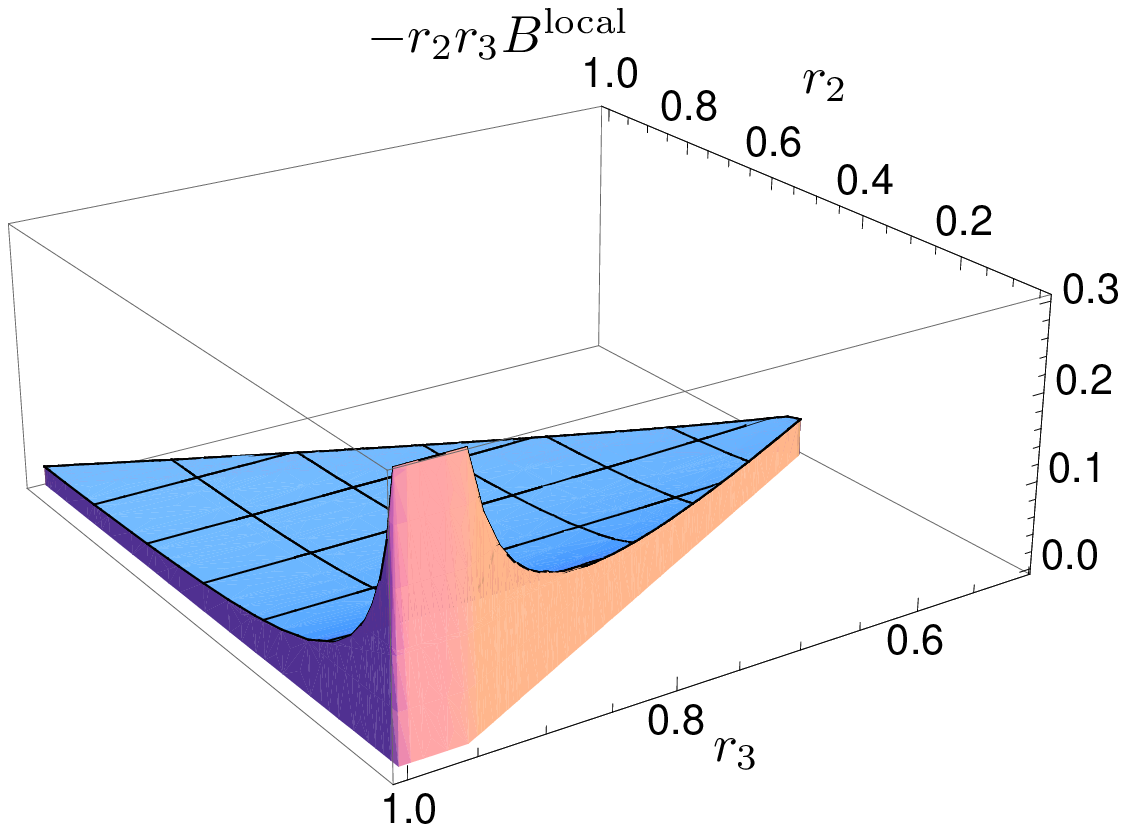}}}
\end{tabular}
\caption{\small The CMB bispectrum on large angular scales induced by primordial non-Gaussianities of
the local form for $f_{\rm NL}^{\rm
  local} = 1$. According to its definition, the bispectrum is
negative for positive $f_{\rm NL}^{\rm
  local}$; thus, we have plotted it with an overall minus sign.}
\label{fig:local}
\end{figure}
\end{center}

A popular shape, usually used in data analysis, is the one obtained when the potential $\phi$ contains a non-linear correction in coordinate space,
\be
\label{local}
\phi (\x) = \phi_g (\x) - f_{\rm NL}^{\rm local} (\phi_g^2 (\x) - \langle \phi_g^2\rangle )\;.
\ee
(We are using the same sign convention for $f_{\rm NL}^{\rm local}$ as Komatsu et al.~\cite{Komatsu:2003fd}.)
In this case, the 3-point function of the gravitational potential $\phi$ is
\be
\label{localFour}
\langle \phi_{\k_1} \phi_{\k_2} \phi_{\k_3} \rangle = (2\pi)^3 \delta (\k_1 + \k_2 + \k_3) ( - 2 f_{\rm NL}^{\rm local} A^2) \left( \frac{1}{k_1^3 k_2^3} + \frac{1}{k_1^3 k_3^3} + \frac{1}{k_2^3 k_3^3} \right) \;.
\ee
If the non-linear correction (\ref{local}) dominates over those computed in the previous section, then $a_{\l}$ can be simply computed using eq.~(\ref{a_l}). 
By taking the ensemble average of the product of three $a_{\l}$ and using eq.~(\ref{localFour}), the CMB bispectrum induced by local non-linear corrections reads
\begin{equation}
B^{\rm local} = - \frac{2  f_{\rm NL}^{\rm
  local}  A^2 }{27 \pi^2}  \left(\frac{1}{l_1^2 l_2^2}+\frac{1}{l_1^2 l_3^2}+\frac{1}{l_2^2 l_3^2}\right) \;.
\end{equation}
Note that by rescaling $l_2$ and $l_3$ we can pull out an overall factor $l_1^{-4}$ and rewrite this bispectrum in terms of the two independent variables $r_2$ and $r_3$,
\begin{equation}
\label{eq:local}
B^{\rm local} = - \frac{2  f_{\rm NL}^{\rm
  local}  A^2 }{27 \pi^2 l_1^4}  \left(\frac{1}{r_2^2}+\frac{1}{r_3^2}+\frac{1}{r_2^2 r_3^2}\right) \;.
\end{equation}
In the following we will always use this trick and plot the bispectrum as a function of $r_2$ and $r_3$ setting $l_1 =1$ and $A=1$. 
The shape corresponding to eq.~(\ref{eq:local}) is plotted in figure~\ref{fig:local}. 

\begin{center}
\begin{figure}[htc]
\begin{tabular}{cc}
\resizebox*{0.48\textwidth}{!}{\includegraphics{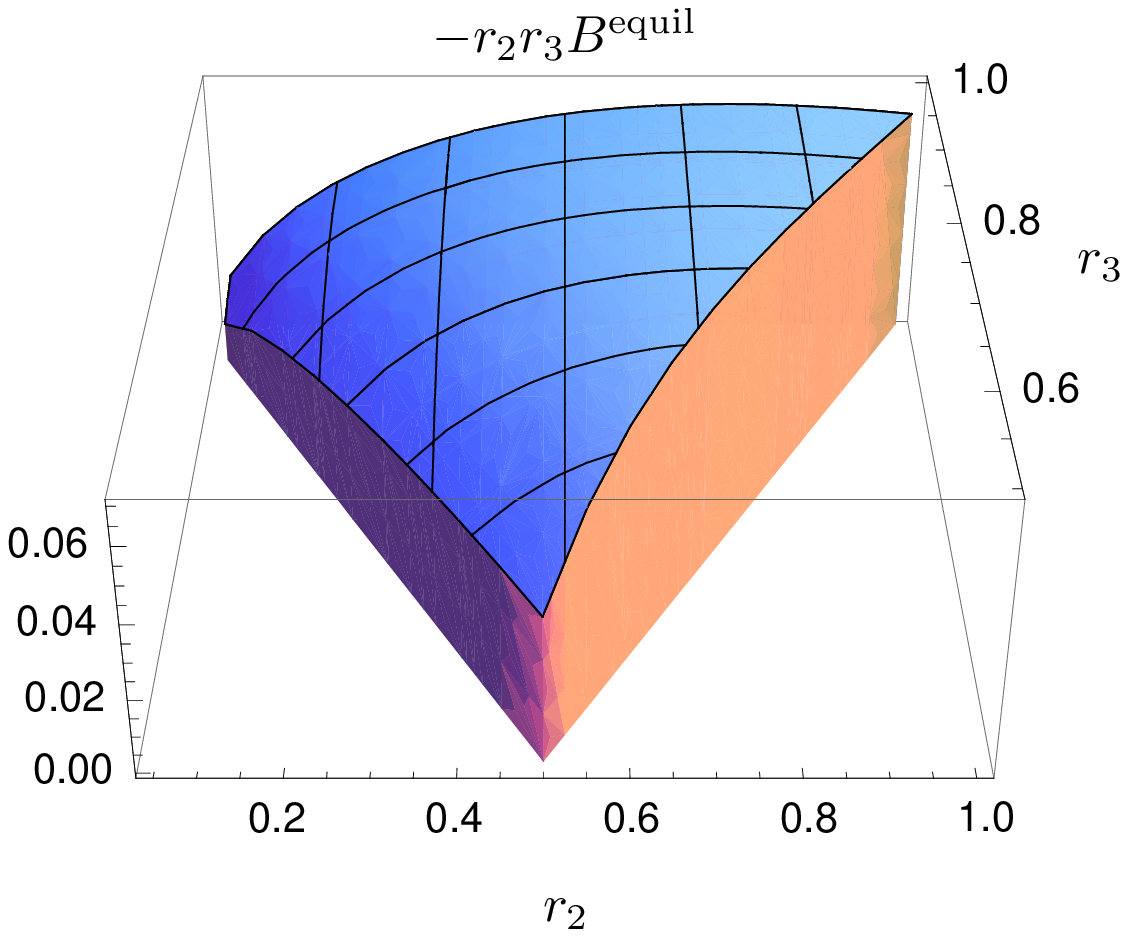}}   &
\raisebox{0.9cm}{\resizebox*{0.48\textwidth}{!}{\includegraphics{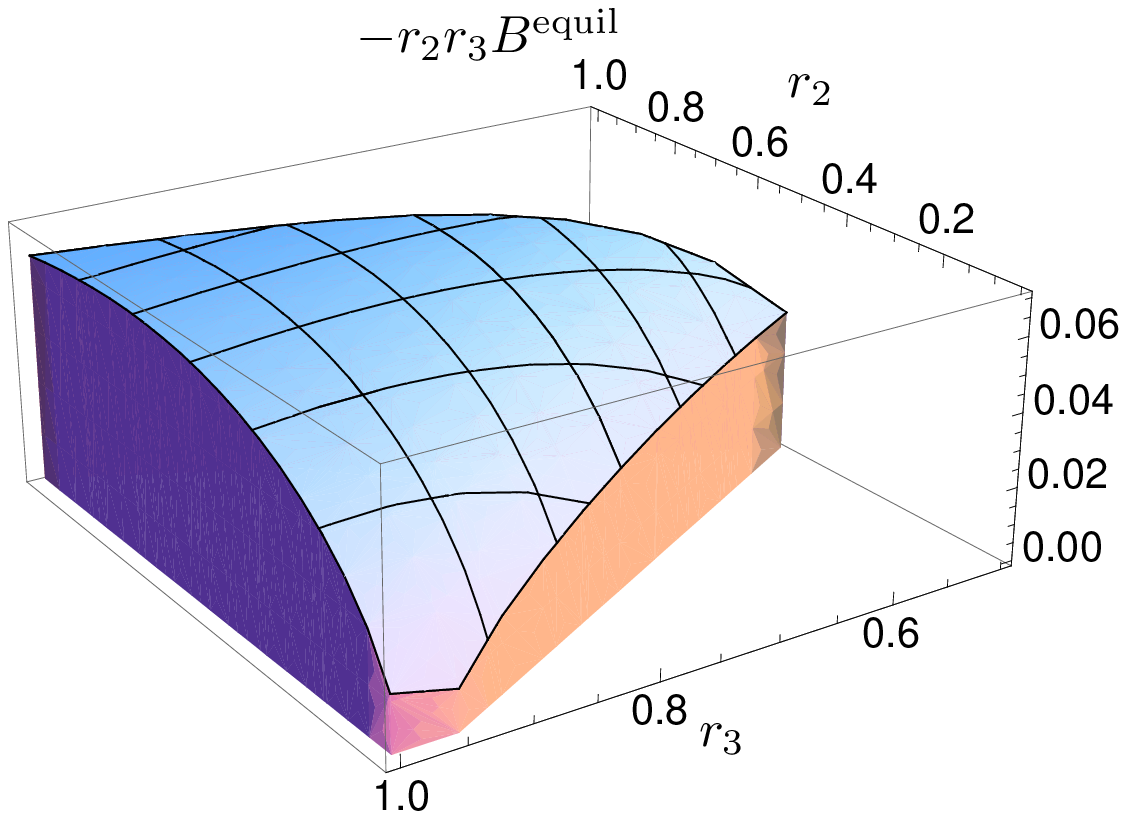}}}
\end{tabular}
\caption{\small The CMB bispectrum on large angular scales induced by primordial non-Gaussianities of
the equilateral form for $f_{\rm NL}^{\rm
  equil} = 1$. According to its definition, the bispectrum is
negative for positive $f_{\rm NL}^{\rm
  equil}$; thus, we have plotted it with an overall minus sign.}
\label{fig:equil}
\end{figure}
\end{center}

\subsection{The equilateral shape}  
Another theoretically motivated shape for the primordial 3-point
function is the so-called equilateral shape, that can be described by \cite{Babich:2004gb}
\be
\langle \phi_{\k_1} \phi_{\k_2} \phi_{\k_3} \rangle = (2\pi)^3 \delta (\k_1 + \k_2 + \k_3) ( - 6 f_{\rm NL}^{\rm equil} A^2) \left( - \frac{1}{ 2 k_1^3 k_2^3}  - \frac{1}{3 k_1^2k_2^2k_3^2} + \frac{1}{k_1 k_2^2 k_3^3} + 5 \ {\rm perms.} \right) \;.
\label{equil_bi}
\ee
Notice that the divergence in the squeezed limit is in this case
milder than for the local shape, due to a cancellation among the
various terms.
We can compute the CMB bispectrum similarly to what is done in
the local case. It is convenient to define 
\be 
y_1 \equiv  {k^\perp_1} (D_e /{l_1}) \;, \qquad y_2 \equiv  {k^\perp_2} (D_e /{l_2})\;. 
\ee
With such a definition, using eq.~(\ref{a_l}) for the $a_{\l}$ and eq.~(\ref{equil_bi}) for the expectation value of three gravitational potentials,  one finally obtains
\begin{equation}
  \begin{split}
\label{Bequil}
B^{\rm equil} =  \frac{2 f_{\rm NL}^{\rm equil}
  A^2}{9 (2\pi)^2 l_1^4} \int_{-\infty}^{+\infty} \mathrm{d} y_1 \mathrm{d} y_2
\left(\frac{1}{2(y_2^2+r_2^2)^{3/2}(y_1^2+r_1^2)^{3/2}}
+ \frac{1}{3(y_1^2+r_1^2)(y_2^2+r_2^2)\big((y_1+y_2)^2+r_3^2\big)} \right. \\ \left.- 
\frac{1}{(y_1^2+r_1^2)^{1/2}(y_2^2+r_2^2)\big((y_1+y_2)^2+r_3^2\big)^{3/2}}
+ 5\,\mathrm{perms.}\right)\;.
  \end{split}
\end{equation}
Here and in the following we sum over all permutations of $(r_1, r_2, r_3)$ and we subsequently set $r_1=1$.
The integrals cannot be done analytically but the result is plotted in figure~\ref{fig:equil}.

From figures \ref{fig:local} and \ref{fig:equil}  we see that the CMB bispectra preserve
in 2d the qualitative features of the primordial 3-point functions:
the signal is peaked on squeezed and equilateral configurations respectively.

%%%%%%%%%%%%%%%%%%%%%%%%%%%%%%%%%%%%%%%%%%%%%%%
%%%%%%%%%%%%%%%%%%%%%%%%%%%%%%%%%%%%%%%%%%
\section{\label{bispectrumcompute}Computing the CMB bispectrum}
%%%%%%%%%%%%%%%%%%%%%%%%%%%%%%%%%%%%%%%%%%%%%%%
%%%%%%%%%%%%%%%%%%%%%%%%%%%%%%%%%%%%%%%%%%

In this section we compute the CMB bispectra due to the different
second-order contributions in eq.~(\ref{CMBfinal}). For comparison, we will use the
two typical primordial shapes, local and equilateral, discussed above. We are only interested in computing the CMB non-Gaussianities generated in the Sachs-Wolfe limit; thus, as already mentioned, we will assume that there is no primordial non-Gaussianity, i.e.~that the curvature perturbation on uniform-density hypersurfaces, $\zeta$, is Gaussian on super-Hubble scales.
Consequently, from eq.~(\ref{zeta_phi}) it follows that $\phi$ is Gaussian.

\subsection{Intrinsic contributions at last scattering}
\label{sec:intrinsic}

Let us start by computing the CMB non-Gaussianity due to the second-order effects in the first line of eq.~(\ref{CMBfinal}), i.e., 
\be
  \frac{\delta T}{{T}}(\n) \supset  \left[    \frac{1}{18}\phi^2
  +\frac{1}{3}\de^{-2}\big((\de_i\phi)^2 -
  3\de^{-2}\de_i\de_j(\de_i\phi\de_j\phi)\big) \right]_e \,. \label{IntrinsicdT}
\ee
This contribution has been first derived in \cite{Bartolo:2004ty} and
its bispectrum and detectability have been studied in
\cite{Liguori:2005rj}.  Note that, although we have dubbed it
``intrinsic'', this contribution is not physically separable from 
the other second-order contributions integrated along the photon path
that we will study below.

The momentum-independent quadratic term, $ \phi_e^2/18$, gives a
contribution to the bispectrum exactly of the local shape, equivalent
to $f_{\rm NL}^{\rm local} = -1/6$ \cite{Bartolo:2004ty}, in eq.~(\ref{eq:local}). 
Its contribution does not vanish in the equilateral limit. We can compare it to an equilateral contribution by evaluating its bispectrum in the equilateral configuration. We find
\be
\frac{B^{\rm -1/6}(1,1,1)}{B^{\rm equil}(1,1,1)} \simeq - 0.24\;,
\ee
where we have evaluated $B^{\rm equil}(1,1,1)$ for $f_{\rm NL}^{\rm
  equil}=1$. We conclude that this contribution is equivalent to
$f^{\rm equil}_{\rm NL} \simeq - 0.24$ in the equilateral limit.

In order to compute the contribution from the momentum-dependent term, we rewrite it as
\be
\frac{1}{3}\de^{-2}(\de_i\phi_e)^2 -
  \de^{-4}\de_i\de_j(\de_i\phi_e \de_j\phi_e) =   \int  \frac{\rmd^3 p_1}{(2 \pi)^3}  \frac{ \rmd^3 p_2}{(2 \pi)^3}
  f^{\rm intr}(\p_1,\p_2) \, \phi_{\p_1}  \phi_{\p_2}  e^{i (\p_1 + \p_2) \cdot \n \; D_e} \;,
\ee
where $f^{\rm intr}(\p_1,\p_2)$ is a kernel defined as
\be
\label{kernel_intr}
  f^{\rm intr}(\p_1,\p_2) \equiv   \frac{1}{3} \frac{\p_1 \cdot \p_2}{(\p_1 + \p_2)^2}
    - \frac{p_1^2 \; p_2^2 + (p_1^2 + p_2^2) (\p_1 \cdot \p_2)
      + (\p_1 \cdot \p_2)^2}{(\p_1 + \p_2)^4} \;.
\ee
Note that this kernel vanishes in the limit of either $p_1$ or $p_2$
going to zero. Thus, we expect this contribution to be suppressed with
respect to the local shape in the squeezed limit.

\begin{figure}[htc]
\begin{center}
\begin{tabular}{cc}
\resizebox*{0.48\textwidth}{!}{\includegraphics{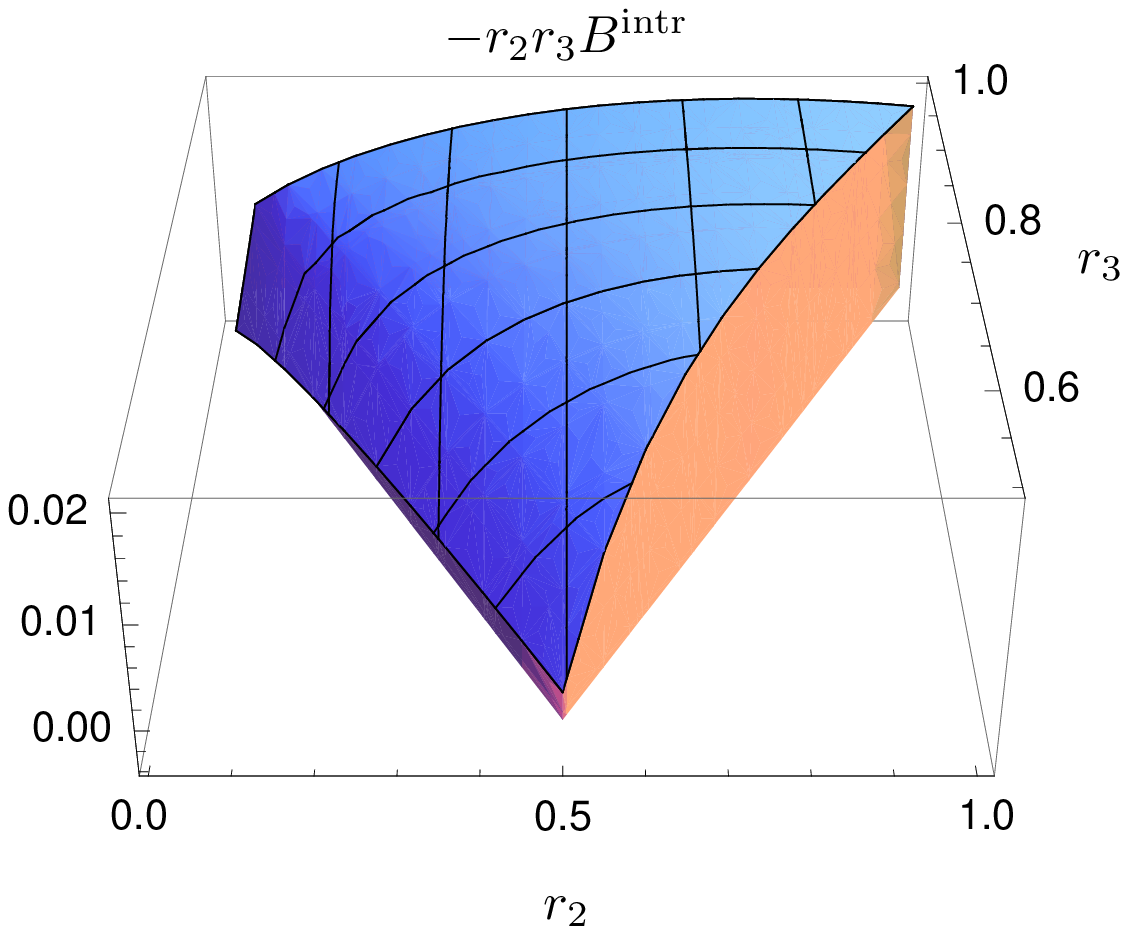}}   &
\raisebox{0.9cm}{\resizebox*{0.48\textwidth}{!}{\includegraphics{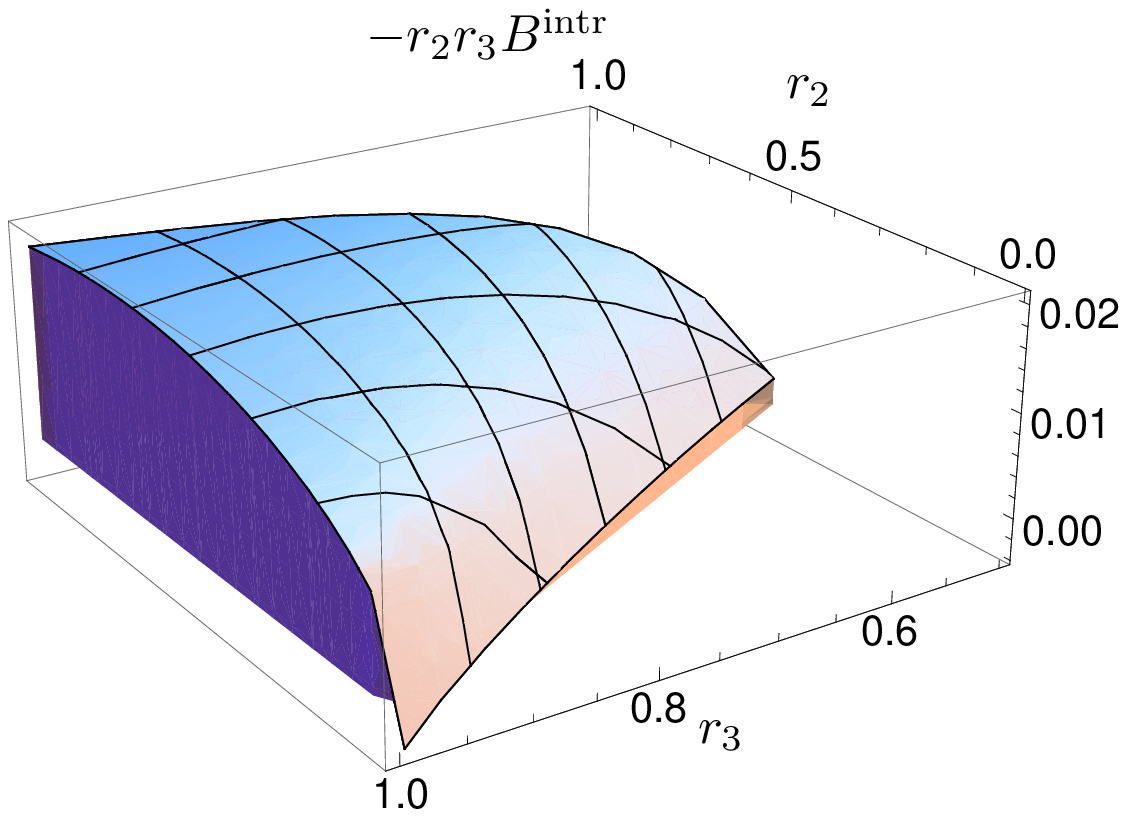}}}
\end{tabular}
\caption{\small The CMB bispectrum induced by the momentum dependent
  intrinsic contribution in eq.~(\ref{IntrinsicdT}).
\label{fig:intr}}
\end{center}
\end{figure}

The Fourier transform in the sky of this contribution is
\be 
a_{\l} =  
  \int \frac{\rmd^3 p_1 }{(2 \pi)^3} \frac{\rmd^3 p_2}{(2 \pi)^3}
  f^{\rm intr}(\p_1,\p_2) \, \phi_{\p_1}  \phi_{\p_2} e^{i (p^\perp_1 + p^\perp_2)  D_e}  (2 \pi)^2\del (\l- (\p_1^{\parallel}+\p_2^{\parallel}) D_e)\;.
\ee
To compute the bispectrum we can contract this contribution, which is quadratic in $\phi$, with the product of two linear Sachs-Wolfe effects, whose $a_{\l}$ are given by eq.~(\ref{a_l}). By doing so, evaluating the 4-point function of $\phi$ using Wick's theorem and the definition of the power spectrum, eq.~(\ref{power_spectrum}), summing over all permutations, and using the definition of the bispectrum, eq.~(\ref{def_bispectrum}), one obtains
\begin{multline}
B^{\rm intr}  = \frac{2 
  A^2}{9 (2\pi)^2 l_1^4} \int_{-\infty}^{+\infty} \mathrm{d} y_1 \mathrm{d} y_2
\Bigg[ \frac{1}{(y_1^2 + r_1^2)^{3/2}(y_2^2 + r_2^2)^{3/2}} \Bigg(\frac{2 y_1 y_2 + r_3^2 - r_1^2 - r_2^2 }{6 \big((y_1+y_2)^2+r_3^2\big)}
\\ - \frac{4 (y_1^2 +r_1^2)(y_2^2 +r_2^2)+2 (y_1^2 +r_1^2+y_2^2 +r_2^2)(2 y_1 y_2 + r_3^2 - r_1^2 - r_2^2) + (2 y_1 y_2 + r_3^2 - r_1^2 - r_2^2)^2 }{4\big((y_1+y_2)^2+r_3^2\big)^2} \Bigg) \\
+ 2\,\mathrm{cyclic} \Bigg] \;.
\label{intr_bis}
\end{multline}
The integrals in the expression above can be integrated numerically. The final result for the bispectrum coming from this contribution is plotted in figure~\ref{fig:intr}. Its contribution is equivalent to $f^{\rm equil}_{\rm NL} \simeq 1.21$.

Notice that this bispectrum is suppressed in the squeezed limit with
respect to the local case in figure \ref{fig:local}. This, as
discussed, is a consequence of the derivatives in
eq.~(\ref{IntrinsicdT}). Notice also that the suppression, in the limit
$r_2 \to 0$, is linear in $r_2$ as there is one derivative acting on
each $\phi$ in eq.~(\ref{IntrinsicdT}). Thus, in the plots (which
include a measure $r_2 r_3$) the function goes to a constant. This
constant depends on the orientation between the long wavelength mode
and the short ones as it is clear from eq.~(\ref{kernel_intr}): indeed,
in the figure we see that the limit $r_2 \to 0$ depends
on the direction from which the limit is approached. Notice that this
behaviour is different from the case of primordial equilateral
non-Gaussianity where there is a suppression going like
$r_2^2$ in the squeezed limit -- $B^{\rm equil}$ diverges logarithmically
for $r_2 \to 0$, see eq.~(\ref{Bequil}) -- so that the plot in
figure \ref{fig:equil} goes to zero.  Indeed, in this limit the 3d kernel
(\ref{equil_bi}) is suppressed by two powers of $k_3$ with respect to the local shape and
this behavior is typical of all equilateral models \cite{Creminelli:2003iq,Babich:2004gb}.

\subsection{Contribution from the Rees-Sciama effect}
\label{sec:RS}
At second-order in the perturbations, the Newtonian and curvature potentials $\Phi$ and $\Psi$ have a constant and a time-dependent part. While the constant part given in the first line of eqs.~(\ref{Phi}) and (\ref{Psi}) dominates on large scales, on sub-Hubble scales one recovers the standard Newtonian limit \cite{Peebles}, i.e.~the two potentials become equal, $\Phi =\Psi$, and grow as the scale factor, $\Phi \propto (aH)^{-2} \propto a$, where we have used $a \propto \tau^2$.
Thus, we expect the photon frequency to be affected by an integrated effect. This is the so-called Rees-Sciama effect \cite{Rees:1968zz}, given by
\be
  \frac{\delta T}{{T}}(\n) \supset     
  \int_{\tau_e}^{\tau_o}\rmd\tau\, \left(\Phi' + \Psi' \right) \,. \label{CMBRS}
\ee
%FV
Its contribution to the CMB bispectrum has already been considered in
\cite{Mollerach:1995sw,Munshi:1995eh} although these analysis were
restricted only to the diagonal terms of the bispectrum. More
generally, the bispectrum from the Rees-Sciama effect has been studied
in \cite{Spergel:1999xn}.\footnote{As there is an error in the
  derivation of eq.~(23) of \cite{Spergel:1999xn}, our results cannot
  be compared with that reference.} 

Symmetrizing over the momenta, we can rewrite the integrand in eq.~(\ref{CMBRS}) as
\be
\Phi' + \Psi' =  \frac{1}{D_e}  \int  \frac{\rmd^3 p_1}{(2 \pi)^3}  \frac{ \rmd^3 p_2}{(2 \pi)^3} 
  f^{\rm RS}(\p_1,\p_2) \, \phi_{\p_1}  \phi_{\p_2}  e^{i (\p_1 + \p_2) \cdot \n D(\tau)} \;,
\ee
where $f^{\rm RS}$ is an explicitly time-dependent kernel derived from eqs.~(\ref{Phi}) and (\ref{Psi}) defined as
\begin{equation}
\label{kernel_RS}
  f^{\rm RS}(\p_1,\p_2) \equiv - \tau D_e  
 \frac{ 4 (\p_1 \cdot \p_2)^2 + 10 p_1^2 p_2^2
    + 7 (p_1^2 + p_2^2) (\p_1 \cdot \p_2) }{21 (\p_1 + \p_2)^2}\;.
\end{equation}
Note that we have multiplied it by $D_e$ to make it dimensionless. 
The Fourier transform in the sky of this contribution is given by
\be 
\label{al_RS}
a_{\l} =  
\int_{\tau_e}^{\tau_o} \frac{d \tau}{D_e}  \int \frac{\rmd^3 p_1 }{(2 \pi)^3} \frac{\rmd^3 p_2}{(2 \pi)^3}
  f^{\rm RS}( \p_1,\p_2) \, \phi_{\p_1}  \phi_{\p_2} e^{i (p^\perp_1 + p^\perp_2)  D (\tau)}  (2 \pi)^2\del (\l- (\p_1^{\parallel}+\p_2^{\parallel}) D(\tau))\;.
\ee

As done for the intrinsic contribution, in order to compute the
bispectrum we need to contract $a_{\l}$ in the above equation with the
product of two linear Sachs-Wolfe effects, whose $a_{\l}$ are given by
eq.~(\ref{a_l}). Note, however, that the Rees-Sciama kernel $f^{\rm RS}$
in eq.~(\ref{kernel_RS}) is higher order in the spatial gradients with
respect to the intrinsic kernel $f^{\rm intr}$ of
eq.~(\ref{kernel_intr}), so that one may think that its contribution
to the bispectrum will be relevant only on short scales. Indeed, since we are correlating the Rees-Sciama
effect with the linear Sachs-Wolfe effect, which takes place at the
last scattering surface, one may naively conclude that its contribution
to the bispectrum is suppressed in the limit of large angles, i.e.~in
the limit where gradients are much smaller than the Hubble rate at
decoupling. 
However, this is not the case. Indeed, the correlation with what
happens at the last scattering surface does not vanish immediately
for $\tau > \tau_e$, but for a given mode $l$, it remains large for $\tau
\lesssim \tau_* \equiv D_e/l$ and after that decays
exponentially. In other words the correlation decays when the distance
from the last scattering surface is of the order of the typical
wavelength. In appendix \ref{app:flat_sky} we explain
better this point with a simple example. Now, since the Rees-Sciama effect grows with $\tau$, the
contribution to the bispectrum will be maximal for $\tau \approx
\tau_* $. Using that $k \sim l/D_e$ one has that the maximal
contribution comes for $k \tau_* \sim 1$ so that the gradients are not
suppressed at $\tau_*$ and one expects the Rees-Sciama contribution to
the bispectrum to be of the same order as one of the intrinsic terms. Notice also that, as for the
intrinsic kernel, also the kernel (\ref{kernel_RS}) vanishes in
the limit of either $p_1$ or $p_2$ going to zero; we thus expect the
Rees-Sciama bispectrum to be suppressed in the squeezed limit with
respect to the local shape.

Let us move to the explicit calculation. It is convenient to define 
\be
x \equiv (\tau - \tau_e)({l_1 }/{D_e})\;. 
\ee
By contracting $a_{\l}$ given by eq.~(\ref{al_RS}) with the product of two linear contributions given by eq.~(\ref{a_l}), using Wick's theorem and the definition of the power spectrum, eq.~(\ref{power_spectrum}),  to rewrite the 4-point function of $\phi$, and summing over all permutations one obtains, by using the variables $y_1$ and $y_2$,
\begin{equation}
  \begin{split}
    B^{\rm RS} = - \frac{2A^2}{189 (2\pi)^2} \frac{1}{l_1^4}
    \int_0^{l_1} \rmd x \, (x+\tau_e(l_1/D_e))
    \int_{-\infty}^{+\infty} \rmd y_1  \rmd y_2 \;
   e^{i (y_1 + y_2) x}  \left[ \frac{1}{(y_1^2 + r_1^2)^{3/2}
    (y_2^2 + r_2^2)^{3/2} } \right.  \\
    \times \left. \left(\frac{3}{2} r_1^2 + \frac{3}{2} r_2^2 + r_3^2
      + 2 y_1 y_2 + \frac{5}{2} (y_1^2 + y_2^2)
      - \frac{5}{2}
      \frac{(y_1^2 - y_2^2 + r_1^2 - r_2^2)^2}{(y_1+y_2)^2 + r_3^2} \right)
    + 2 \ {\rm cyclic} \right]\;.
\label{RS2}
  \end{split}
\end{equation}
Actually the result of the calculation is not proportional to
$(2\pi)^2 \delta(\l_1 + \l_2 + \l_3)$ as in the definition of
eq.~(\ref{def_bispectrum}), but to $(2\pi)^2 \delta\left((\l_1 + \l_2)
  \frac{\tau_o-\tau}{\tau_o-\tau_e} + \l_3\right)$ and permutations, 
as a consequence of the fact that we are correlating effects at
different conformal times $\tau$. 
This is a bit surprising as the delta function $\delta(\l_1 + \l_2 + \l_3)$ is just a
consequence of translational invariance. However, the discussion above
implies  that the bispectrum is exponentially suppressed when the
triangle in Fourier space does not close, i.e.~when
$\frac{\tau_o-\tau}{\tau_o-\tau_e} l_3 \sim 1$. This can be checked
explicitly in the expression (\ref{RS2}). In appendix
\ref{app:flat_sky} we discuss a simple example in which this issue is
made more transparent. 

\begin{figure}[htc]
\begin{center}
\begin{tabular}{cc}
\resizebox*{0.48\textwidth}{!}{\includegraphics{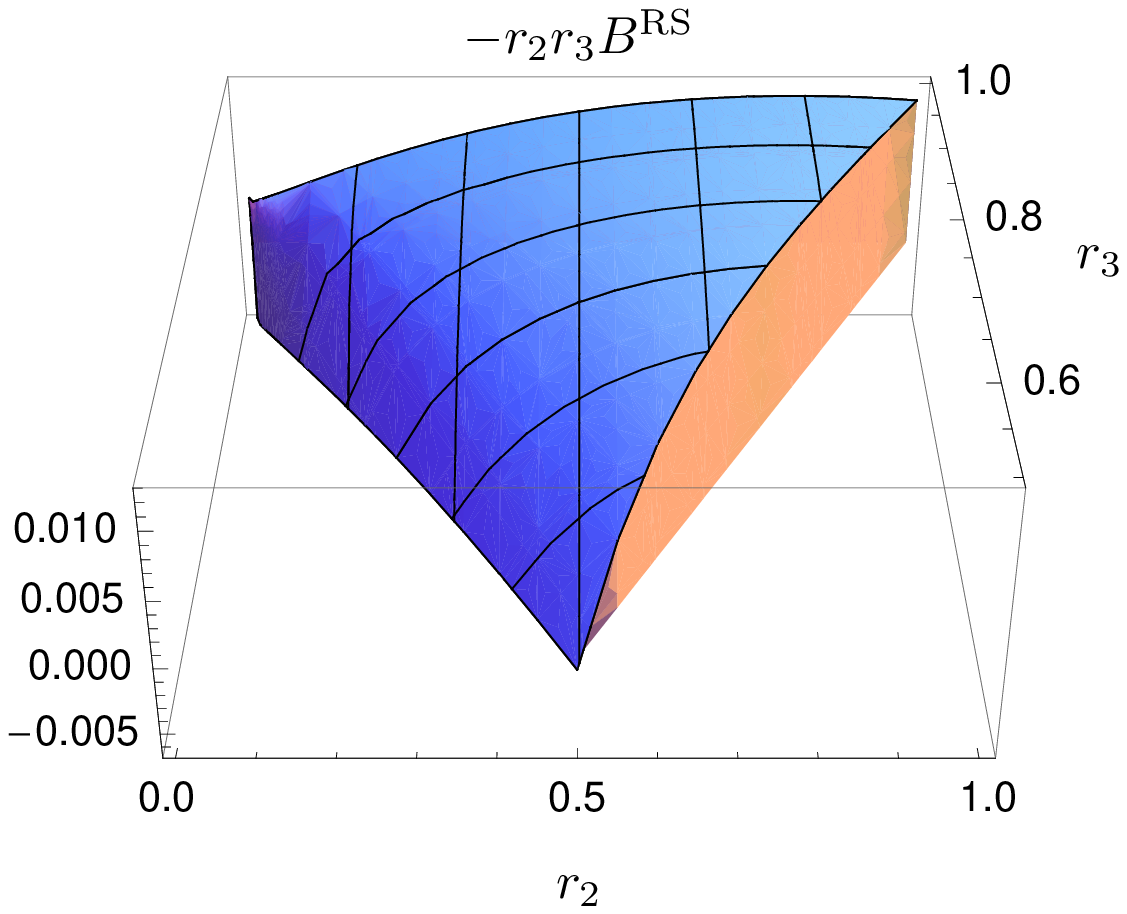}}   &
\raisebox{0.9cm}{\resizebox*{0.48\textwidth}{!}{\includegraphics{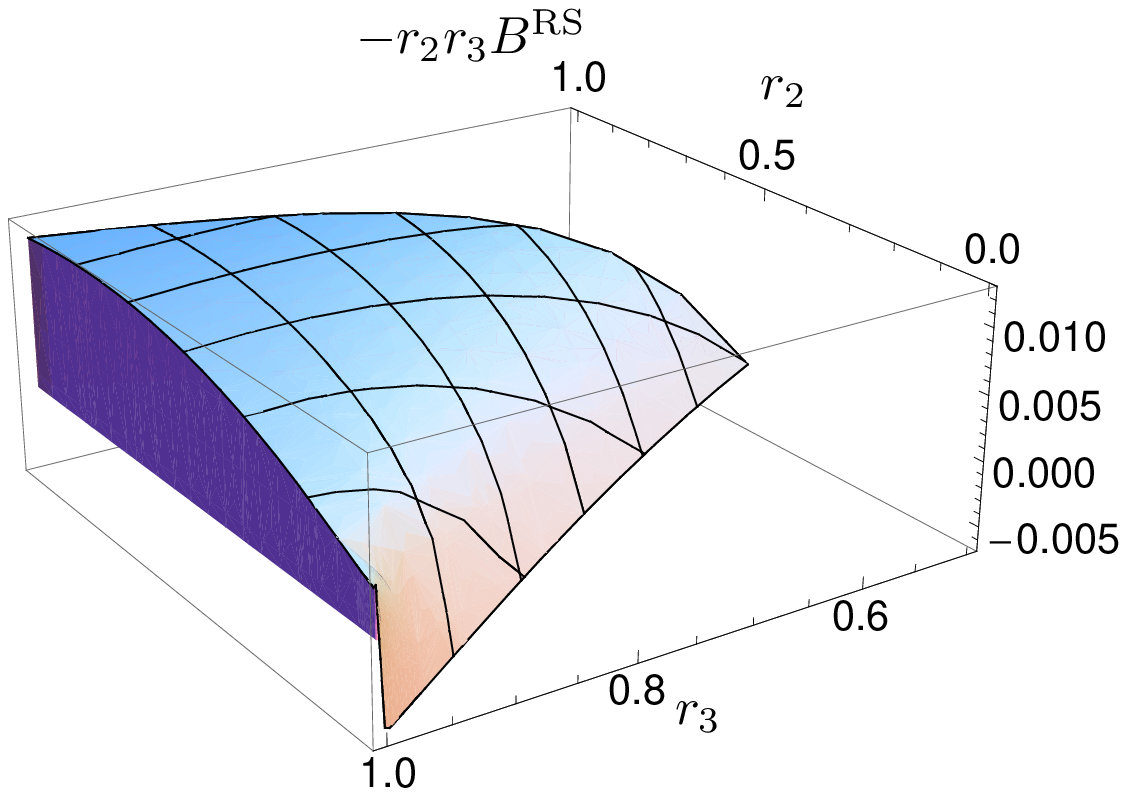}}}
\end{tabular}
\caption{\small The CMB bispectrum induced by the Rees-Sciama effect, eq.~(\ref{CMBRS}).}
\label{fig:RS}
\end{center}
\end{figure}

The above integrals are particularly challenging even
numerically. However, some simplifications can be made. Since the
integrand is exponentially suppressed for $x \gg 1$ by the rapid
oscillations of $e^{i (y_1 +y_2) x}$, one can push the upper limit of
the integral in $x$ to $\infty$. Another simplification consists in
neglecting $\tau_e(l_1/D_e)$ in the first integral of eq.~\eqref{RS2}, which is
justified by the fact that we consider only modes well outside the
Hubble radius at recombination and thus $ \tau_e (l_1 /D_e) \sim \tau_e k \ll 1$. With these approximations eq.~(\ref{RS2}) can be rewritten as 
\begin{equation}
  \begin{split}
    B^{\rm RS} = - \frac{2A^2}{189 (2\pi)^2} \frac{1}{l_1^4}
    \int_0^{\infty} \rmd x \, x
    \int_{-\infty}^{+\infty} \rmd y_1  \rmd y_2 \;
   e^{i (y_1 + y_2) x}  \left[ \frac{1}{(y_1^2 + r_1^2)^{3/2}
    (y_2^2 + r_2^2)^{3/2} } \right.  \\
    \times \left. \left(\frac{3}{2} r_1^2 + \frac{3}{2} r_2^2 + r_3^2
      + 2 y_1 y_2 + \frac{5}{2} (y_1^2 + y_2^2)
      - \frac{5}{2}
      \frac{(y_1^2 - y_2^2 + r_1^2 - r_2^2)^2}{(y_1+y_2)^2 + r_3^2} \right)
    + 2 \ {\rm cyclic} \right]\;.
\label{RS}
  \end{split}
\end{equation}
We see that the bispectrum induced by the Rees-Sciama effect goes as
$l^{-4}$ and it is parametrically similar to the intrinsic contribution discussed in the previous section.
The analytical and numerical study of this expression is postponed to
appendix \ref{app:RS}. The final result for the bispectrum is given in
figure \ref{fig:RS}.

As for the intrinsic contribution (\ref{intr_bis}), in the squeezed
limit $r_2 \to 0$ the Rees-Sciama bispectrum is suppressed when compared with
the local shape by $r_2$, with a coefficient which depends
on the angle. We show this analytically in appendix~\ref{app:RS}. 
By comparing the Rees-Sciama bispectrum to the equilateral contribution, as we did for the intrinsic one, we find that the Rees-Sciama contribution is equivalent to $f_{\rm NL}^{\rm equil} \simeq 0.74$.

% Comparison with Spergel & Goldberg

\subsection{Integrated vector contribution}

At second order, the non-diagonal part of the metric $g_{0i}\equiv a^2
\omega_i$ becomes non-vanishing and time dependent on sub-Hubble
scales. Similarly to the time-dependent part of the gravitational
potentials, it induces an integrated effect on the photon redshift,
given in eq.~(\ref{CMBfinal}) by 
\be
\label{CMBV}
  \frac{\delta T}{{T}}(\n) \supset   
  \int_{\tau_e}^{\tau_o}  \rmd\tau \;
  \omega_i'\hat{n}^i \,.
\ee
As $\omega_i$ is transverse we refer to this effect as the integrated vector contribution.
As for the Rees-Sciama, to compute the bispectrum we need to correlate
this integrated effect with the intrinsic temperature fluctuation at
last scattering. Even though this effect is suppressed at last
scattering, when modes are still out of the Hubble radius, it will give us a contribution to $f_{\rm NL}$ of order unity, similarly to what happens for the Rees-Sciama effect.

From eq.~(\ref{omega}) we can rewrite the integrand as
\be
\omega_i' \hat n^i =  \frac{1}{D_e}  \int  \frac{\rmd^3 p_1}{(2 \pi)^3}  \frac{ \rmd^3 p_2}{(2 \pi)^3} 
  f^{\rm V}(\p_1,\p_2) \, \phi_{\p_1}  \phi_{\p_2}  e^{i (\p_1 + \p_2) \cdot \n D(\tau)} \;,
\ee
where $f^{\rm V}$ is a kernel defined as
\begin{equation}
  f^{\rm V}(\p_1,\p_2) = 
  - \frac{2i D_e}{3} 
  \left[ \frac{p_1^2 (\n \cdot \p_2)  + p_2^2 ( \n \cdot \p_1)}{(\p_1 +\p_2)^2}
    - \n \cdot (\p_1 + \p_2) 
    \frac{2 p_1^2 p_2^2 +(p_1^2 +p_2^2) (\p_1 \cdot \p_2) }{(\p_1 +\p_2)^4} \right]\;.
\label{fvector}
\end{equation}

Note that the second term in the kernel (\ref{fvector}) is proportional to $\hat n \cdot (\p_1+ \p_2)$. Thus, it is 
a time total derivative which 
can be trivially integrated in $\tau$ in eq.~(\ref{CMBV}). 
Therefore we have another term evaluated at last scattering, analogous to
the intrinsic contributions studied in section~\ref{sec:intrinsic}, of
the form 
\be
\label{Vbound}
  \frac{\delta T}{{T}}(\n) \supset  \frac43 \left[
    \partial^{-4} \partial_j (\partial^2\phi \partial_j \phi)\right]_e \,. 
\ee
This shows clearly that there is nothing really intrinsic about the
contributions discussed in section~\ref{sec:intrinsic}: the splitting
among the various effects is gauge dependent and only the total sum
has a well defined gauge invariant meaning.

One can then split the rest of the kernel orthogonally to and along the line of sight. Indeed, decomposing $\n$ into the parts orthogonal and parallel to the line of sight as $\n = (\m, 1)$, the first term in eq.~(\ref{fvector}) can be rewritten as
\begin{equation}
  - \frac{2i D_e}{3}  \left[ \frac{\m \cdot  ( \p^\parallel_2 p_1^2   +  \p^\parallel_1 p_2^2  )}{(\p_1 +\p_2)^2} +
   \frac{p_1^2 p^\perp_2  + p_2^2  p^\perp_1 }{(\p_1 + \p_2)^2}
 \right]\;.
\label{vecint}
\end{equation}
The first term of this expression is proportional to $\m$. Thus, it is
higher order in $1/l$ with respect to the second term and therefore
negligible in the flat-sky approximation. 
Thus, the Fourier transform on the sky of the contribution
(\ref{vecint}) can be approximated with
\be 
\label{al_V}
a_{\l} =  
- \frac{2i D_e}{3} \int_{\tau_e}^{\tau_o} \frac{d \tau}{D_e}  \int \frac{\rmd^3 p_1 }{(2 \pi)^3} \frac{\rmd^3 p_2}{(2 \pi)^3} 
   \,\frac{p_1^2 p^\perp_2  + p_2^2  p^\perp_1 }{(\p_1 + \p_2)^2}
\, \phi_{\p_1}  \phi_{\p_2} e^{i (p^\perp_1 + p^\perp_2)  D (\tau)}  (2 \pi)^2\del (\l- (\p_1^{\parallel}+\p_2^{\parallel}) D(\tau))\;.
\ee

\begin{figure}[htc]
\begin{center}
\begin{tabular}{cc}
\resizebox*{0.48\textwidth}{!}{\includegraphics{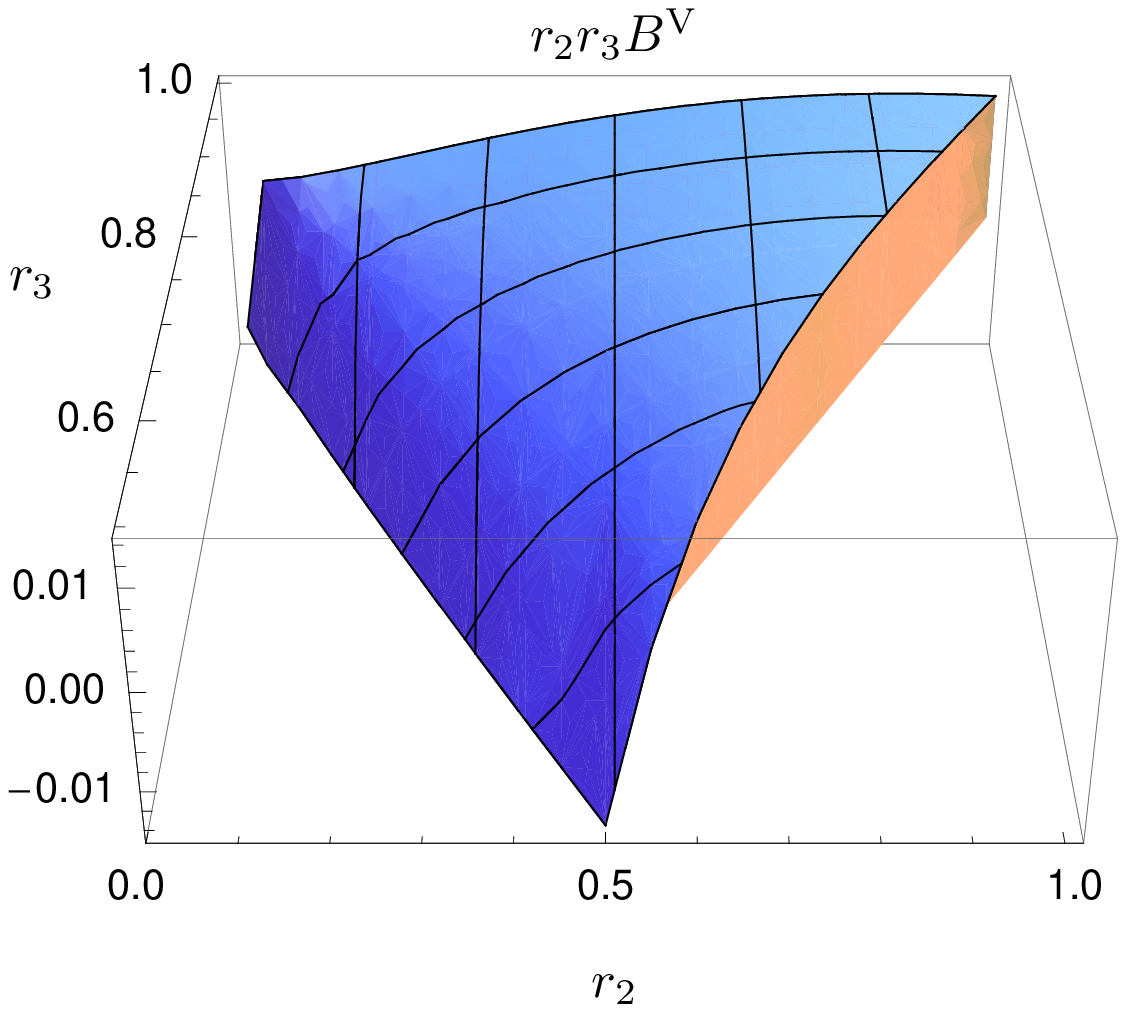}}   &
\raisebox{0.9cm}{\resizebox*{0.48\textwidth}{!}{\includegraphics{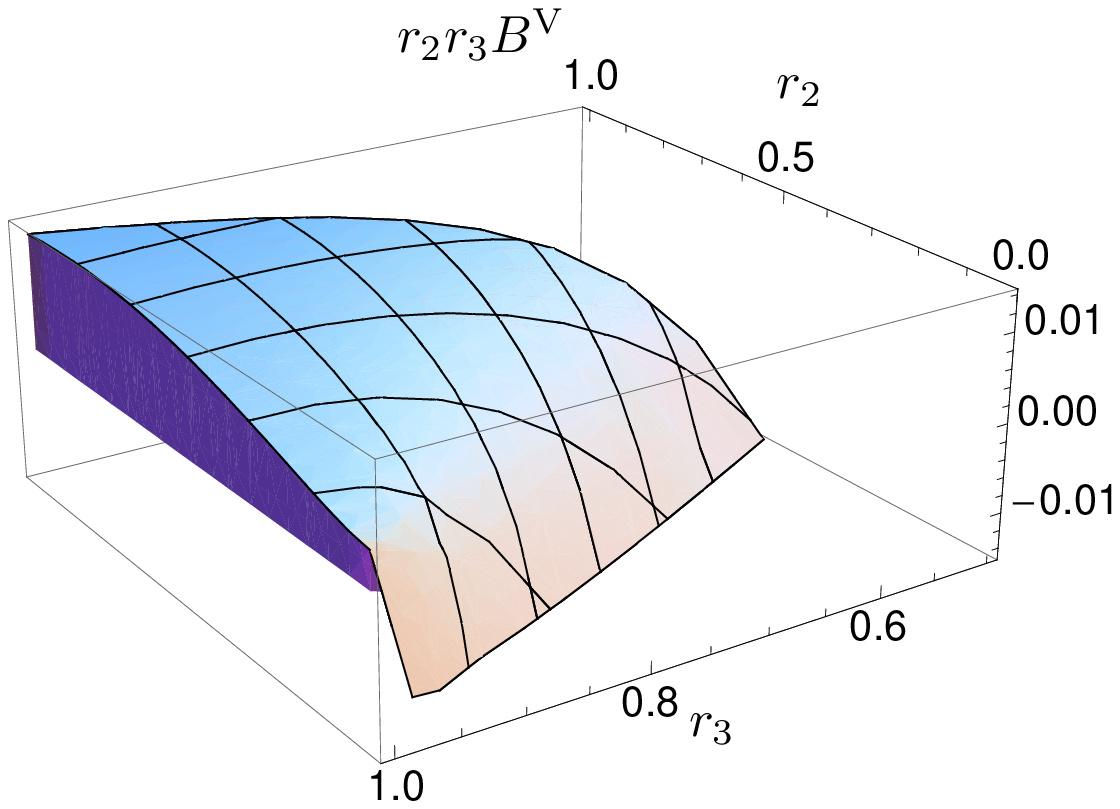}}}
\end{tabular}
\caption{\small The CMB bispectrum induced by the vector contribution, eq.~(\ref{CMBV}).}
\label{fig:V}
\end{center}
\end{figure}

Proceeding as in the case of
the intrinsic and Rees-Sciama contributions, the total contribution from
vectors can be written, using the variables $y_1$ and $y_2$, as
\begin{multline}
  B^{\rm V}= \frac{4A^2}{27 (2\pi)^2
    l_1^4}\int_{-\infty}^{\infty}\rmd y_1
   \rmd y_2\; 
  \Bigg[ \frac{1}{(y_1^2 + r_1^2)^{3/2}(y_2^2 +
  r_2^2)^{3/2}} \\ \times \bigg( \frac{ (y_1^2+r_1^2)\big( 2 y_2(y_1+y_2)-(r_1^2-r_2^2-r_3^2)\big) + (y_2^2+r_2^2)\big(2 y_1(y_1+y_2)-(r_2^2-r_1^2-r_3^2)\big)}{2 \big((y_1+y_2)^2+r_3^2\big)^{2} } \\
+i \int_0^{\infty} \rmd x \, e^{i (y_1 +y_2) x}  \frac{(y_1^2 +
  r_1^2)y_2 + (y_2^2 + r_2^2)y_1}{(y_1+y_2)^2+r_3^2} \bigg) + 2 \ {\rm cyclic} \Bigg] \;.
  \label{V_bis}
  \end{multline}
The first piece, which is not integrated in $x$, comes from
eq.~(\ref{Vbound}), while the other term describes the contribution
integrated along the line of sight. The integral over time can be dealt with as in the Rees-Sciama case:
see appendix \ref{app:RS}.
The final result for this bispectrum is given in figure
\ref{fig:V}. Again, the result is suppressed with respect to the local
shape in the squeezed limit because the kernel (\ref{fvector})
vanishes when either $p_1$ or $p_2$ go to zero. The behaviour in this
limit is qualitatively the same as in the Rees-Sciama case. This
vector contribution is equivalent to $f_{\rm NL}^{\rm equil} \simeq
-0.84$ in the equilateral configuration.

\subsection{Integrated tensor contribution}

At second order, even in the absence of primordial gravitational waves, the part of the spatial metric
not proportional to the identity matrix, $a^2 \gamma_{ij}$, is non-vanishing and time dependent on sub-Hubble scales. Thus, it induces an integrated effect given by
\be
  \frac{\delta T}{{T}}(\n) \supset    -
  \int_{\tau_e}^{\tau_o} \rmd \tau \; \frac{1}{2}\gamma_{ij}'\hat{n}^i\hat{n}^j\,, \label{tensdT}
\ee
which we expect to contribute to the bispectrum similarly to what
happens for the vectors.
As $\gamma_{ij}$ is transverse and traceless, we refer to this effect as the tensor contribution.
From eq.~(\ref{gamma}) and using eqs.~(\ref{expansion2}) and (\ref{Theta}) to rewrite the transverse traceless projector, the integrand is
\be
- \frac{1}{2}\gamma_{ij}'\hat{n}^i\hat{n}^j  =  \frac{1}{D_e}  \int  \frac{\rmd^3 p_1}{(2 \pi)^3}  \frac{ \rmd^3 p_2}{(2 \pi)^3} 
  f^{\rm T}(\p_1,\p_2) \, \phi_{\p_1}  \phi_{\p_2}  e^{i (\p_1 + \p_2) \cdot \n \, D(\tau)} \;,
\ee
where the kernel $ f^{\rm T}$ is defined as
\begin{equation}
  \begin{split}
  f^{\rm T}(\p_1,\p_2) =  - j_2 (|\p_1 +\p_2| \tau)
  \frac{5 D_e}{ \tau}  
  \left[  \frac{ (\p_1 \cdot \p_2)^2 - p_1^2 p_2^2   }{ (\p_1 +\p_2)^4}
    \left( 1 + \frac{(\n \cdot (\p_1 +\p_2))^2}{(\p_1 +\p_2)^2} \right)  \right. \\
 \left. + \frac{2 p_1^2 (\n \cdot \p_2 )^2 + 2 p_2^2 (\n \cdot \p_1)^2
  - 4 (\p_1 \cdot \p_2) (\n \cdot \p_1 ) (\n \cdot \p_2 )}{(\p_1 +\p_2)^4} \right]\;,
\end{split}
\end{equation}
and $j_2$ is a spherical Bessel function that appears from taking the time derivative of $\gamma_{ij}$,
\be
\left(\frac{ j_1 (k \tau)}{k \tau} \right)' = - \frac{ j_2 (k \tau)}{\tau} \;.
\ee

\begin{figure}[htc]
\begin{center}
\begin{tabular}{cc}
\resizebox*{0.48\textwidth}{!}{\includegraphics{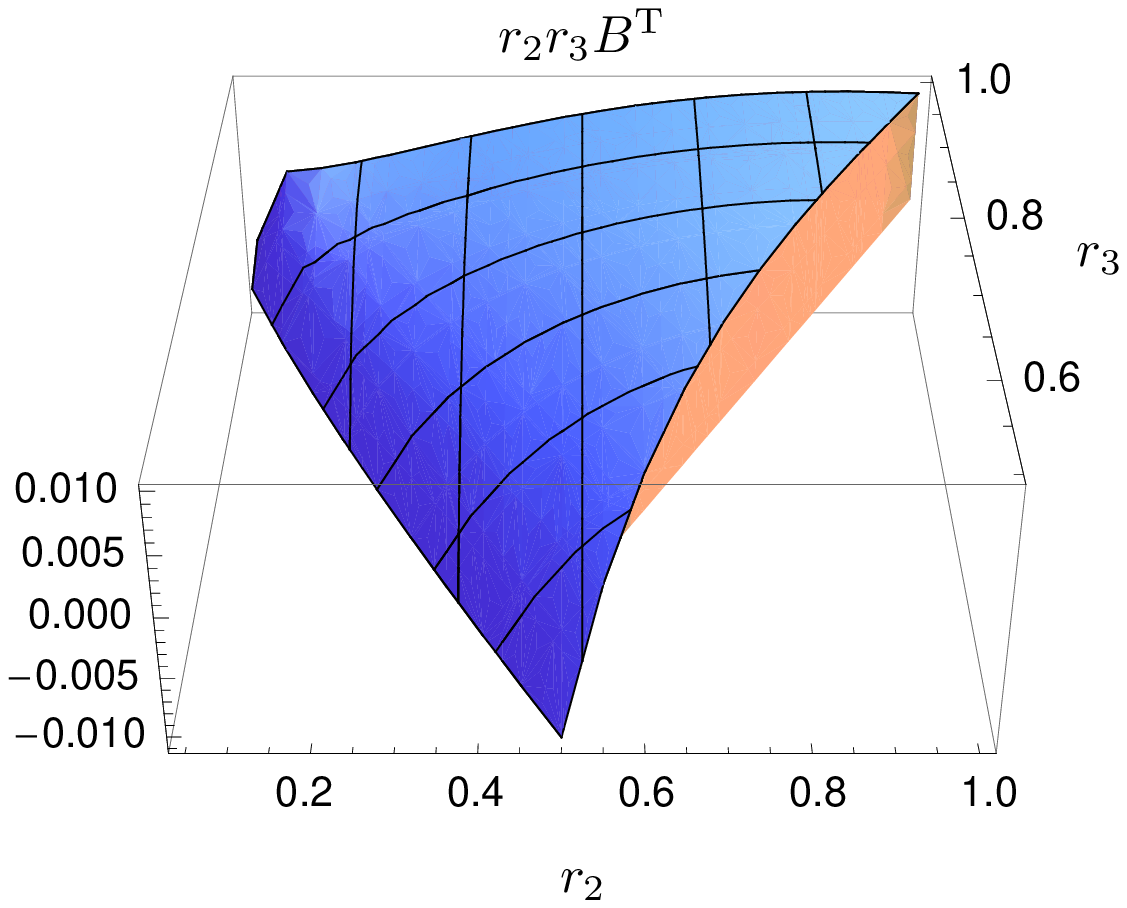}}   &
\raisebox{0.9cm}{\resizebox*{0.48\textwidth}{!}{\includegraphics{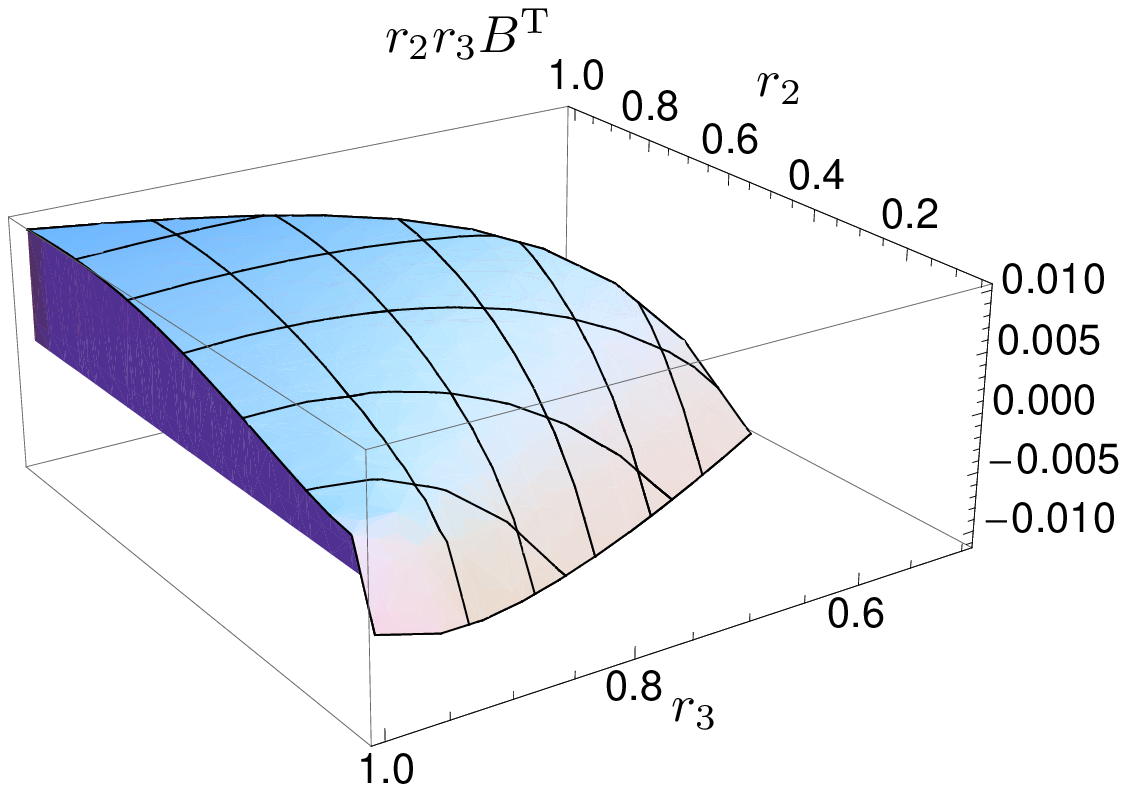}}}
\end{tabular}
\caption{\small The CMB bispectrum induced by the tensor contribution, eq.~(\ref{tensdT}).}
\label{fig:T}
\end{center}
\end{figure}

As we did for the vector kernel $f^{\rm V}$, $f^{\rm T}$ can be decomposed into a part parallel and orthogonal to the sky. The parallel part is higher order in $1/l$ and thus negligible in the flat-sky approximation. Thus, the kernel can be approximated as
\begin{equation}\label{kernelT}
  \begin{split}
  f^{\rm T}(\p_1,\p_2) \simeq  - j_2 (|\p_1 +\p_2| \tau)
  \frac{5 D_e}{ \tau}  
  \left[  \frac{ (\p_1 \cdot \p_2)^2 - p_1^2 p_2^2   }{(\p_1 +\p_2)^4}
    \left( 1 + \frac{(p^\perp_1 +p^\perp_2)^2}{(\p_1 +\p_2)^2} \right)  \right. \\
 \left. + \frac{2 p_1^2 (p^\perp_2 )^2 + 2 p_2^2 ( p^\perp_1)^2
  - 4 (\p_1 \cdot \p_2) p^\perp_1 p^\perp_2 }{(\p_1 +\p_2)^4} \right]\;.
\end{split}
\end{equation}
The Fourier transform on the sky of this contribution is given by
\be 
\label{al_T}
a_{\l} =  
\int_{\tau_e}^{\tau_o} \frac{\rmd \tau}{D_e}  \int \frac{\rmd^3 p_1 }{(2 \pi)^3} \frac{\rmd^3 p_2}{(2 \pi)^3}
  f^{\rm T}( \p_1,\p_2) \, \phi_{\p_1}  \phi_{\p_2} e^{i (p^\perp_1 + p^\perp_2)  D (\tau)}  (2 \pi)^2\del (\l- (\p_1^{\parallel}+\p_2^{\parallel}) D(\tau))\;.
\ee
With this simplification the time integral can be analytically computed and yields, expressing it in terms of the variables $x$, $y_1$ and $y_2$,
\be
\int_0^{\infty} \!\! d x \, \frac{ j_2 (\sqrt{(y_1+y_2)^2+r_3^2} \,x)}{ x} e^{i (y_1 + y_2) x} =  \frac{(2 r_3^2 - (y_1 + y_2)^2) }{6 ((y_1 + y_2)^2 + r_3^2)} - \frac{
  (y_1 + y_2) r_3^2 \coth^{-1} \left(\frac{\sqrt{
   (y_1 + y_2)^2+r_3^2 }}{y_1 + y_2} \right) }{2 ((y_1 + y_2)^2+r_3^2)^{3/2}} \;,
\ee
plus an imaginary term odd under $(y_1, y_2)  \to (-y_1,-y_2)$ which does not contribute to the integral. This gives for the bispectrum
\begin{equation}
\begin{split}
  &B^{\rm T} =  \frac{10 A^2}{9 (2 \pi)^2 l_1^4}  
  \int_{-\infty}^{+\infty}\rmd y_1 \rmd y_2  \left[ \frac{(2 r_3^2 - (y_1 + y_2)^2) }{6 ((y_1 + y_2)^2 + r_3^2)} - \frac{
  (y_1 + y_2) r_3^2 \coth^{-1} \left(\frac{\sqrt{
   (y_1 + y_2)^2+r_3^2 }}{y_1 + y_2} \right) }{2 ((y_1 + y_2)^2+r_3^2)^{3/2}} \right] \\
   & \times \frac{1}{(y_1^2 + r_1^2)^{3/2} (y_2^2 + r_2^2)^{3/2}
    ((y_1+y_2)^2 + r_3^2)^{2}} \\
   & \times  \Bigg[
\frac14   \left(1 + \frac{(y_1 + y_2)^2}{(y_1 + y_2)^2 + r_3^2} \right)
    \left( 4 (y_1^2 + r_1^2) (y_2^2 + r_2^2) 
      - (2 y_1 y_2 + r_3^2 - r_1^2 - r_2^2)^2 \right) \\
   &\qquad - 2 \left( y_2^2 (y_1^2 + r_1^2) + y_1^2 (y_2^2 + r_2^2)
      - y_1 y_2 (2 y_1 y_2 + r_3^2 - r_1^2 - r_2^2) \right) \Bigg]+ 2 \ {\rm cyclic} \Bigg\},
\label{T_bis}
\end{split}
\end{equation}
and the final result is plotted in figure \ref{fig:T}. Again, given
that the kernel (\ref{kernelT}) goes to zero when either $p_1$ or
$p_2$ go to zero, this shape is suppressed with respect to
the local one in the squeezed limit. From figure \ref{fig:T} we see
that the integrated tensor
contribution is qualitatively similar to the intrinsic kernel, Rees-Sciama
and vector contributions discussed previously.
This contribution is equivalent to $f_{\rm NL}^{\rm equil} \simeq
-0.61$ for an equilateral configuration.

\subsection{Lensing}
\label{sec:lensing}

The deflection angle of a light ray as it propagates from the last
scattering surface to us is given by eq.~(\ref{alpha2}) (for a review of lensing effects on the CMB
see \cite{Lewis:2006fu}). For convenience we reproduce it here,  
\begin{equation}
\vec{\alpha} = -2\int_{\tau_e}^{\tau_o}\rmd\tau\;\frac{\tau - \tau_e}{\tau_o - \tau_e} \, \vec \nabla_{\parallel}\phi\,.
\label{alpha}
\end{equation}
The geometrical weight $(\tau - \tau_e)/(\tau_o - \tau_e)$ tells us
that the effect is suppressed close to the last scattering
surface. For this reason, usually the main contribution to the 3-point function
due to lensing comes from the correlation of the photon deflection with the
ISW \cite{Seljak:1998nu,Goldberg:1999xm}. This effect is absent in our case
as we are studying a universe with only matter. However, there is
still the correlation of the intrinsic temperature fluctuation at
last scattering with the lensing contribution given by
\begin{align}
  \frac{\delta T}{{T}}(\n) \supset    
  \frac{1}{3} \vec \alpha  \cdot \vec \nabla_{\n} \phi_e  \,. \label{lensingdT}
\end{align}
Similarly to the other integrated effects also this will give an effective $f_{\rm NL} \sim 1$.

Let us compute the contribution to the bispectrum. 
Inserting the deviation angle (\ref{alpha}) into eq.~(\ref{lensingdT})
and using $\vec \nabla_{\hat n} = D_e \vec \nabla_\parallel $, the lensing contribution to the temperature fluctuation can be written as
\be
  \frac{\delta T}{{T}}(\n) \supset    
   \frac{2}{3} \int_{\tau_e}^{\tau_o}\rmd\tau \, (\tau - \tau_e) \int  \frac{\rmd^3 p_1}{(2 \pi)^3}  \frac{ \rmd^3 p_2}{(2 \pi)^3}  (\p_1^\parallel  \cdot  \p_2^\parallel)  \phi_{\p_1} \phi_{\p_2} e^{i \p_1 \cdot \n D(\tau)} e^{ i \p_2 \cdot \n D_e}   \,. \label{lensingdT2}
\ee
Taking the Fourier transform on the sky yields
\be
a_{\l} = \frac{2}{3} \int_{\tau_e}^{\tau_o}\rmd\tau \, (\tau - \tau_e) \int  \frac{\rmd^3 p_1}{(2 \pi)^3}  \frac{ \rmd^3 p_2}{(2 \pi)^3}  (\p_1^\parallel  \cdot  \p_2^\parallel)  \phi_{\p_1} \phi_{\p_2} e^{i p^\perp_1  D(\tau)} e^{ i p^\perp_2 D_e}  (2 \pi)^2  \delta(\l - \p_1^\parallel D(\tau) - \p_2^\parallel D_e) \,.
\ee

\begin{figure}[htc]
\begin{center}
\begin{tabular}{cc}
\resizebox*{0.48\textwidth}{!}{\includegraphics{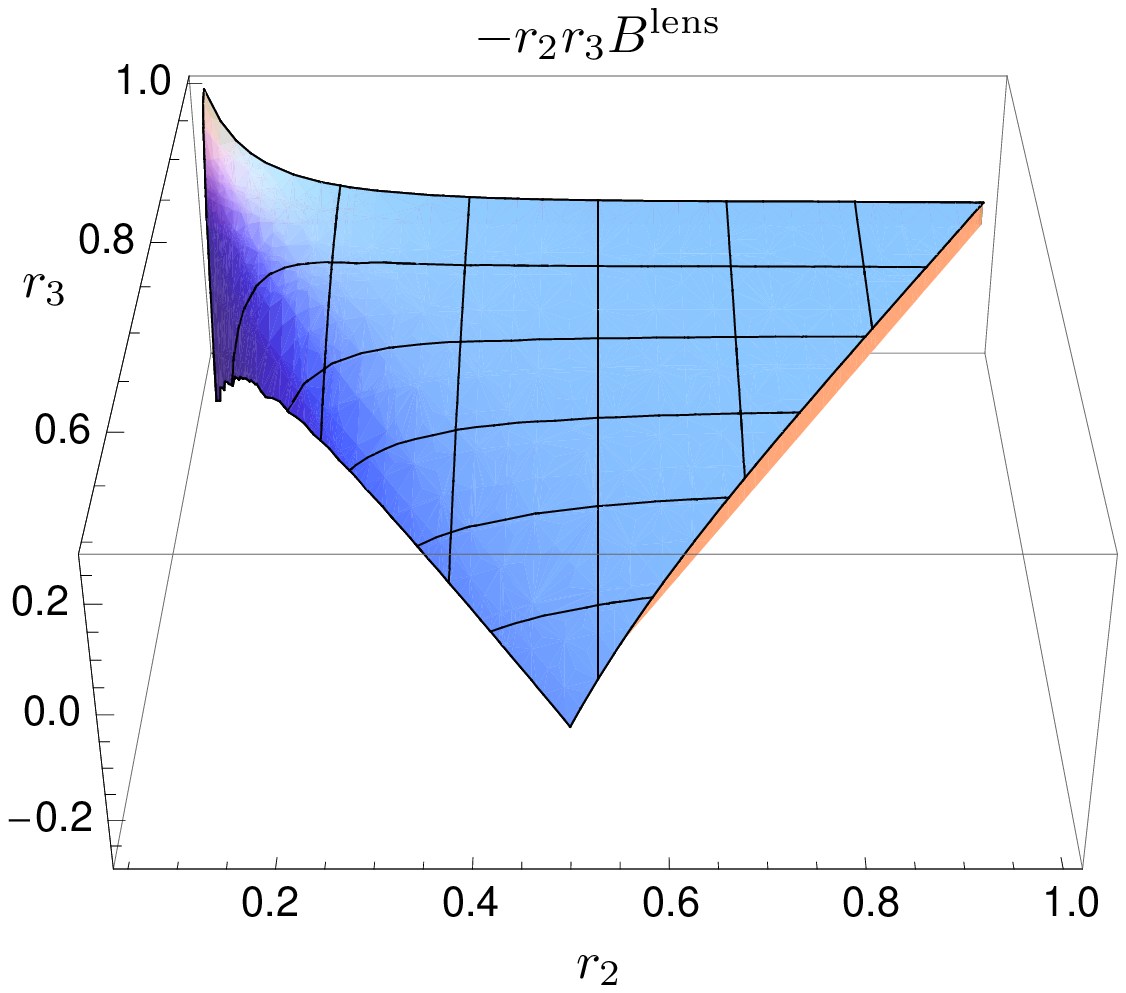}}   &
\raisebox{0.9cm}{\resizebox*{0.48\textwidth}{!}{\includegraphics{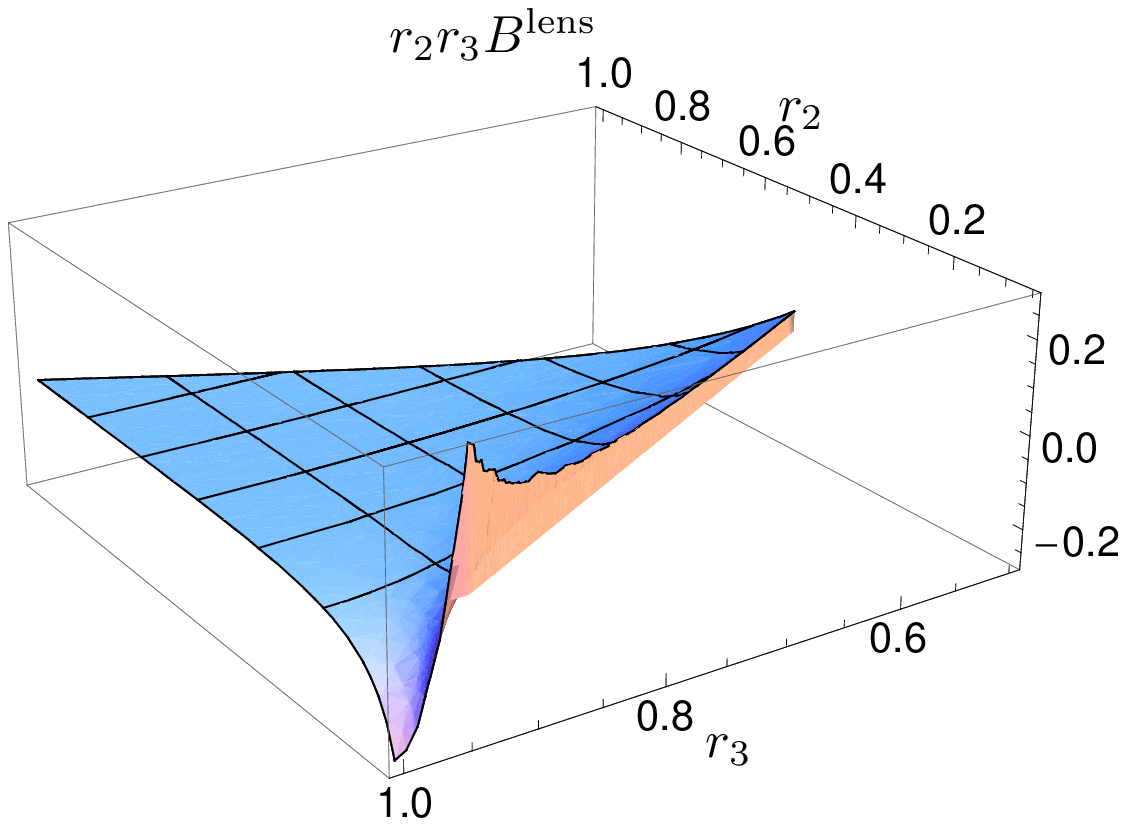}}}
\end{tabular}
\caption{\small The CMB bispectrum induced by the lensing contribution, eq.~(\ref{lensingdT}).}
\label{fig:lens}
\end{center}
\end{figure}

As usual, we can compute the bispectrum by correlating this effect with the intrinsic temperature at last scattering. By doing so, we obtain
\be
B^{\rm lens} = -\frac{A^2}{27 (2\pi)^2 l_1^4} \int_0^{\infty}\rmd x\; x \int_{-\infty}^\infty \rmd y_1 \rmd y_2 \, e^{i y_1 x } \frac{r_3^2 - r_1^2 - r_2^2}{(y_1^2 + r_1^2)^{3/2} (y_2^2 + r_2^2)^{3/2}}\\ + 5 \ \rm{perms.}\,.
\ee
Note that here one must sum over all permutations of $(r_1,r_2,r_3)$, including the anticyclic ones. The  integrals above can be computed analytically, yielding
\begin{equation}
\label{eq:lensexplicit}
B^{\rm lens} = \frac{8 A^2}{27 (2\pi)^2 l_1^4}\frac{r_3^2 - r_1^2 - r_2^2}{r_1^4 r_2^2} + 5 \ \rm{perms.}
\end{equation}
This result is plotted in figure~\ref{fig:lens}. Alternatively, this equation can be written as 
\be
\label{eq:lensexplicit2}
B^{\rm lens} =  \frac{16 A^2}{27 (2\pi)^2 } \frac{\l_1 \cdot \l_2 }{l_1^4 l_2^2} + 5 \ \rm{perms.}
\ee
Another method to derive the lensing CMB bispectrum is through the
lensing potential $\psi$ defined as (see for example \cite{Lewis:2006fu}) 
\begin{equation}
\psi(\hat{n})\equiv -2 \int_{\tau_e}^{\tau_o} \rmd \tau\; \frac{\tau -
  \tau_e}{(\tau_o -\tau_e)(\tau_o -\tau)}\, \phi\left(\hat{n}(\tau_o
  -\tau), \tau\right) \;.
\end{equation}
The deflection angle (\ref{alpha}) is obtained by taking the flat-sky
gradient of this expression $\vec\alpha=\vec\nabla_{\hat{n}}\psi$. The
correlation between the temperature at the last scattering surface and
the lensing potential is given by\footnote{As explained in \cite{Lewis:2006fu}, the divergence of the lensing potential at $\tau_o$ affects only the monopole, which can always be subtracted.}
\begin{equation}
\langle \psi_{\vec l_1} a_{\vec l_2}\rangle=  -\frac{8 \pi A}{3
  D_e^2 l_2} \int^{\tau_o}_{\tau_e} \rmd \tau
\,\frac{(\tau-\tau_e)^2}{(\tau_o-\tau)}\; K_1(l_2 (\tau-\tau_e)/D_e) \;\delta (\vec l_1+
  \vec l_2({\tau_o-\tau})/{D_e} )\;.
\label{lens}
\end{equation}
The temperature fluctuation is localized at $\tau_e$ while the lensing
becomes more and more important at later times. It is easy to see that
the correlation is maximal at $\tau_* \sim \tau_o/l$, similarly to
what was discussed for all integrated effects in section \ref{sec:RS} (see also
appendix \ref{app:flat_sky}).

%  The correlated effects peak at different scales: SW peaks at $\tau\sim\tau_e$, while lensing is maximal at $\tau_o$.  As it is clear from the exercise of Appendix A, the cross-correlation peaks at some intermediate conformal time between $\tau_e$ and $\tau_o$. Now, let us consider the limit where $\tau$ is very close, but not equal, to $\tau_o$ in Eq.(\ref{lens}), to maximize the lensing contribution.  In this limit, the cross-correlation is dominated by long wavelength lensing mode, i.e. $\ell_1\approx 0$. This is important when considering the squeezed limit of the lensing bispectrum in Sect. 4.5. The lensing mode will play the role of the background wave. 
 
As the integral is dominated by $\tau \ll \tau_0$, we can approximate
the $\delta$ function with $\delta(\vec l_1 + \vec l_2)$ to get
\begin{equation}
\langle \psi_{\vec l_1} a_{\vec l_2}\rangle= (2\pi)^2
\delta(\vec l_1 + \vec l_2) C^{T,\psi}_{l_1}\;, \quad\qquad
C^{T,\psi}_{l}=-\frac{4\, A}{3 \pi l^4}. 
\label{cross}
\end{equation}
The bispectrum can be written as \cite{Zaldarriaga:2000ud,Hu:2000ee}
\begin{equation}
 B^{\rm lens}= - \vec l_1\cdot \vec l_2 \left( C_{l_1}
   C^{T,\psi}_{l_2} + C_{l_2} C^{T,\psi}_{l_1}  \right) + \textrm{2 
   perms} \;,
\end{equation}
which coincides with eq.~(\ref{eq:lensexplicit2}). 

%%%%%%%%%%%%%%%%%%%%%%%%%%%%%%%%%%%%%%%%%%%%%%%%%%%%%%%%%%%%%%%%%%%%%%%%%%%%%%%%%%%%
Note that the expression in eq.~(\ref{eq:lensexplicit2}) diverges in the squeezed limit. However, the form of
the divergence depends on the direction one approaches the limit. One can compare the
expression resulting from taking $r_2 \rightarrow 0$ with the local
form. This gives a contribution equivalent to $f^{\rm local}_{\rm NL} = - \cos(2 \,\theta)$ where $\theta$ is the angle between $\vec{l}_2$ and $\vec{l}_1$ when one takes the limit. In the equilateral configuration, the lensing gives a sizable contribution, equivalent to $f^{\rm equil}_{\rm NL} \simeq 2.87$.

It is possible to recover the lensing 3-point function in the squeezed limit in
another way, which is physically more transparent and can be easily
generalized to the case when the short wavelength modes are inside the
horizon at recombination. 
We are going to calculate the 3-point function by first taking
the long wavelength mode fixed and then studying its lensing effect on the short
scale 2-point function.\footnote{It is easy to argue that the leading
  contribution in the squeezed limit is obtained when the lensing mode
  is of long wavelength. Indeed, lensing is effective far from the last
  scattering surface, but as we get far from it the correlation
  with the temperature fluctuation rapidly decreases. The loss of
  correlation happens at $\tau_\ast \sim \tau_o/l$, i.e.~it is faster
  at high $l$, that is why the squeezed limit is
  dominated by a long lensing wave.} At the end we average over the long wavelength
mode.\footnote{This discussion is inspired by the derivation of the
  consistency relation for the squeezed limit of the primordial 3-point function
  in single field inflation
  \cite{Maldacena:2002vr,Creminelli:2004yq,Cheung:2007sv}. In
  particular we will parallel the explicit derivation done in sec.~2 of \cite{Cheung:2007sv}.}
Consider the 2-point correlation function of the temperature fluctuations in two different directions $\n_1$ and $\n_2$. In the presence of a long wavelength mode the real space 2-point function is lensed 
\begin{equation}
\left\langle \frac{\delta T}{T}(\n_1) \frac{\delta T}{T}(\n_2)\right\rangle_{\rm lens} =\left\langle \frac{\delta T}{T} \frac{\delta T}{T} \right\rangle
[\n_1 + \vec\alpha(\n_1) 
- \n_2 - \vec\alpha(\n_2)] \;,
\end{equation}  
where we used the fact that the unlensed 2-point function just depends
on the distance between the points. Obviously there is no effect if
the two lensing angles are the same: the 2-point function is just
translated. Expanding at first order and defining by $\vec m_1$ and
$\vec m_2$ the components of $\n_1$ and $\n_2$ parallel to the (flat) sky we have
\begin{equation}
\left\langle \frac{\delta T}{T}(\n_1) \frac{\delta T}{T}(\n_2)\right\rangle_{\rm lens} =\left\langle \frac{\delta T}{T} \frac{\delta T}{T} \right\rangle
[\m_1 - \m_2] + \nabla_i \left\langle \frac{\delta T}{T}
  \frac{\delta T}{T} \right\rangle \nabla_j \alpha_i \left[\frac{\m_1+\m_2}{2}\right] \cdot (\m_1 - \m_2)_j \;.
\end{equation} 
By assumption the lensing wave is of long wavelength so that we can
evaluate the gradient of the lensing angle at the midpoint $(\m_1
+ \m_2)/2$. If we call $\m \equiv \m_1- \m_2 $, we have
\begin{equation}
\left\langle \frac{\delta T}{T}(\n_1) \frac{\delta T}{T}(\n_2)\right\rangle_{\rm lens} =\left\langle \frac{\delta T}{T} \frac{\delta T}{T} \right\rangle
[m] + \frac{\rmd}{\rmd \log m} \left\langle \frac{\delta T}{T}
  \frac{\delta T}{T} \right\rangle [m] \;\frac{m_j}{m}
\frac{m_i}{m} \;\nabla_j \alpha_i \left[\frac{\vec m_1+\vec m_2}{2}\right]  \;.
\end{equation}
We can now Fourier transform to $\vec l_1$ and $\vec l_2$. The result can be
expressed in terms of $\vec l_S = (\vec l_1 - \vec l_2)/2$ and $\vec
l_L = \vec l_1 + \vec l_2$, where $_L$ and $_S$ stand for long and
short wavelength,
\begin{equation}
\langle a_{\vec l_1} a_{\vec
l_2} \rangle_{\rm lens} =C_{l_S} +i l_{L j} \alpha_i(\vec l_L) \int \rmd^2 m  \frac{\rmd}{\rmd \log m} \left\langle \frac{\delta T}{T}
  \frac{\delta T}{T} \right\rangle [m] \;\frac{m_j}{m}
\frac{m_i}{m} \;e^{-i \vec l_S \vec m} \;.
\end{equation}
The 3-point function is obtained multiplying the above expression by $\delta T/T$ of the
long wavelength mode and averaging,
\begin{equation}
\langle a_{\vec l_1} a_{\vec l_2} a_{\vec l_3} \rangle = (2 \pi)^2
\delta(\vec l_1 + \vec l_2 + \vec l_3) \cdot i l_{L j} \langle \frac{\delta
T}{T} \alpha_i \rangle'(l_L) \int \rmd^2 m  \frac{\rmd}{\rmd \log m} \left\langle \frac{\delta T}{T}
  \frac{\delta T}{T} \right\rangle [m] \;\frac{m_j}{m}
\frac{m_i}{m} \;e^{-i \vec l_S \vec m}  \;.
\label{lens_cons}
\end{equation}
The prime in the correlation between lensing and the temperature means
that we have to remove the momentum conservation factor $(2 \pi)^2 \delta$, which has been
factored out.

Let us evaluate the integral over $\vec m$, which describes the effect of
lensing on the 2-point function. One may na\"ively think that for a scale
invariant 2-point function, which is the case that we are studying in this
paper, the effect of lensing
vanishes. Indeed, the calculations above are very similar to the
ones leading to the consistency relation for the squeezed limit of the
primordial 3-point function
\cite{Maldacena:2002vr,Creminelli:2004yq,Cheung:2007sv}. In that case,
however, the integral over $\vec m$ does not contain the angular weight
$m^jm^i/m^2$. Without this terms the integral vanishes for a scale
invariant spectrum: indeed the 2-point function in real space is a
logarithm of the distance, so that its log-derivative is a constant. The Fourier transform
of a constant is $\delta(\vec l_S)$ which vanishes for any non-zero
$\vec l_S$. 

The situation is different in the presence of the angular
weight $m^jm^i/m^2$. To be more explicit, let us introduce a scale dependence in the 2-point function and evaluate the integral in eq.~(\ref{lens_cons}) for a power spectrum of the form
$C_l = C \cdot l^{-2+(n_s-1)}$, which corresponds to a 2-point
function going as $m^{-(n_s-1)}$, to see that the result does not
vanish for $n_s \to 1$. The integral can be written as
\begin{multline}
\int \rmd^2 m  \frac{\rmd}{\rmd \log m} \left\langle \frac{\delta T}{T}
  \frac{\delta T}{T} \right\rangle [m] \;\frac{m_j}{m}
\frac{m_i}{m} \;e^{-i \vec l_S \vec m} =
-(n_s-1) \frac{\partial_{l_i} \partial_{l_j}}{\nabla^2} \int \rmd^2 m \left\langle \frac{\delta T}{T}
  \frac{\delta T}{T} \right\rangle \;e^{-i \vec l_S \vec m} = \\
= -(n_s-1) \frac{\partial_{l_i} \partial_{l_j}}{\nabla^2} C \cdot
l^{-2+(n_s-1)} = -(n_s-1) \; \partial_{l_i} \partial_{l_j}  C \cdot
\frac{l^{n_s-1}}{(n_s-1)^2} = - C \cdot l^{-2+(n_s-1)}
\left[\delta_{ij} +(n_s-3) \frac{l_i l_j}{l^2} \right] \;.
\end{multline}
We see that the result does not vanish for $n_s=1$. What vanishes for
$n_s=1$ is the trace of this tensor. This means that for a scale
invariant spectrum, the isotropic rescaling due to lensing does not
contribute to the 3-point function. This makes sense in light of the
discussion above: for the isotropic part there is no angular weight so
that everything works as for the consistency relation for the squeezed limit of the
primordial 3-point function
\cite{Maldacena:2002vr,Creminelli:2004yq,Cheung:2007sv}.
On the other hand, the anisotropic case is similar to what happens
when one calculates the primordial 3-point function of a graviton and
two scalar modes, in the limit when the graviton wavelength becomes
very long. The gravitational wave induces an anisotropic rescaling of
the scalar 2-point function and the result does not vanish for a scale
invariant spectrum \cite{Maldacena:2002vr}. An analogous effect is found when computing the contribution 
to the scalar trispectrum from graviton exchange \cite{Seery:2008ax}. In the limit where the graviton wavelength is very long, the non-isotropic rescaling induces a correlation between a pair  of scalar 2-point functions. This effect has the same spin-2 angular dependence as the lensing.

Let us go back to eq.~(\ref{lens_cons}). In our case the normalization of the spectrum is given by $C = A/ (9
\pi)$, so that the expression of the 3-point function in the squeezed
limit gives
\begin{equation}
\langle a_{\vec l_1} a_{\vec l_2} a_{\vec l_3} \rangle = (2 \pi)^2
\delta(\vec l_1 + \vec l_2 + \vec l_3) \cdot i l_{L j} \langle \frac{\delta
T}{T} \alpha_i \rangle'(l_L) \left(- \frac{A}{9 \pi}\right) \frac{1}{l_S^2}
\left(\delta_{ij}- 2 \frac{l_{Si} l_{Sj}}{l_S^2}\right)\;.
\end{equation}
The correlation between the temperature and the deflection angle is
given by 
\begin{multline}
\langle \frac{\delta
T}{T} \alpha_i \rangle'(l_L) = - \frac23 \frac{1}{D_e^2} \int \frac{\rmd k_\perp}{2 \pi}
\int_{\tau_e}^{\tau_o} \rmd \tau \; \frac{\tau-\tau_e}{D_e} \frac{i l_{L i}}{D_e} \frac{A}{(k_\perp^2 +
  l_L^2/D_e^2)^{3/2}} e^{i k_\perp (\tau -\tau_e)} = - \frac{1}{3
  \pi} \cdot \frac{4 A \; i l_{L i}}{l_L^4} \;.
\end{multline}
Thus we have 
\begin{equation}
B^{\rm lens} = - \frac{4 A^2}{27 \pi^2}
\frac{l_{Li} l_{Lj}}{l_L^4 l_S^2} \left(\delta_{ij}- 2 \frac{l_{Si} l_{Sj}}{l_S^2}\right)\;.
\end{equation}
In the limit $\vec l_2 \to 0$, the explicit expression (\ref{eq:lensexplicit2}) gives, taking into account the permutation $l_1 \leftrightarrow l_3$,
\begin{equation}
B^{\rm lens} = \frac{4 A^2}{27 \pi^2}
\frac{1}{l_2^4} \left[\frac{\vec l_1 \cdot \vec l_2}{l_1^2}-
  \frac{\vec l_2 \cdot (\vec l_1+\vec l_2)}{(\vec l_1+ \vec
    l_2)^2}\right] \simeq - \frac{4 A^2}{27 \pi^2} \frac{l_{2i} l_{2j}}{l_2^4}
 \frac{\rmd}{\rmd l_{1j}} \frac{l_{1i}}{l_1^2} \;,
\end{equation}
which coincides with the expression above.

% Comparison with Challinor & Lewis for lensing expression (Filippo)

%%%%%%%%%%%%%%%%%%%%%%%%%%%%%%%%%%%%%%%%%%%%%%%
%%%%%%%%%%%%%%%%%%%%%%%%%%%%%%%%%%%%%%%%%%%%%%%
\section{\label{bisfinal}The total CMB bispectrum}
%%%%%%%%%%%%%%%%%%%%%%%%%%%%%%%%%%%%%%%%%%%%%%%
%%%%%%%%%%%%%%%%%%%%%%%%%%%%%%%%%%%%%%%%%%%%%%%

\begin{center}
\begin{figure}[htc]
\begin{tabular}{cc}
\resizebox*{0.48\textwidth}{!}{\includegraphics{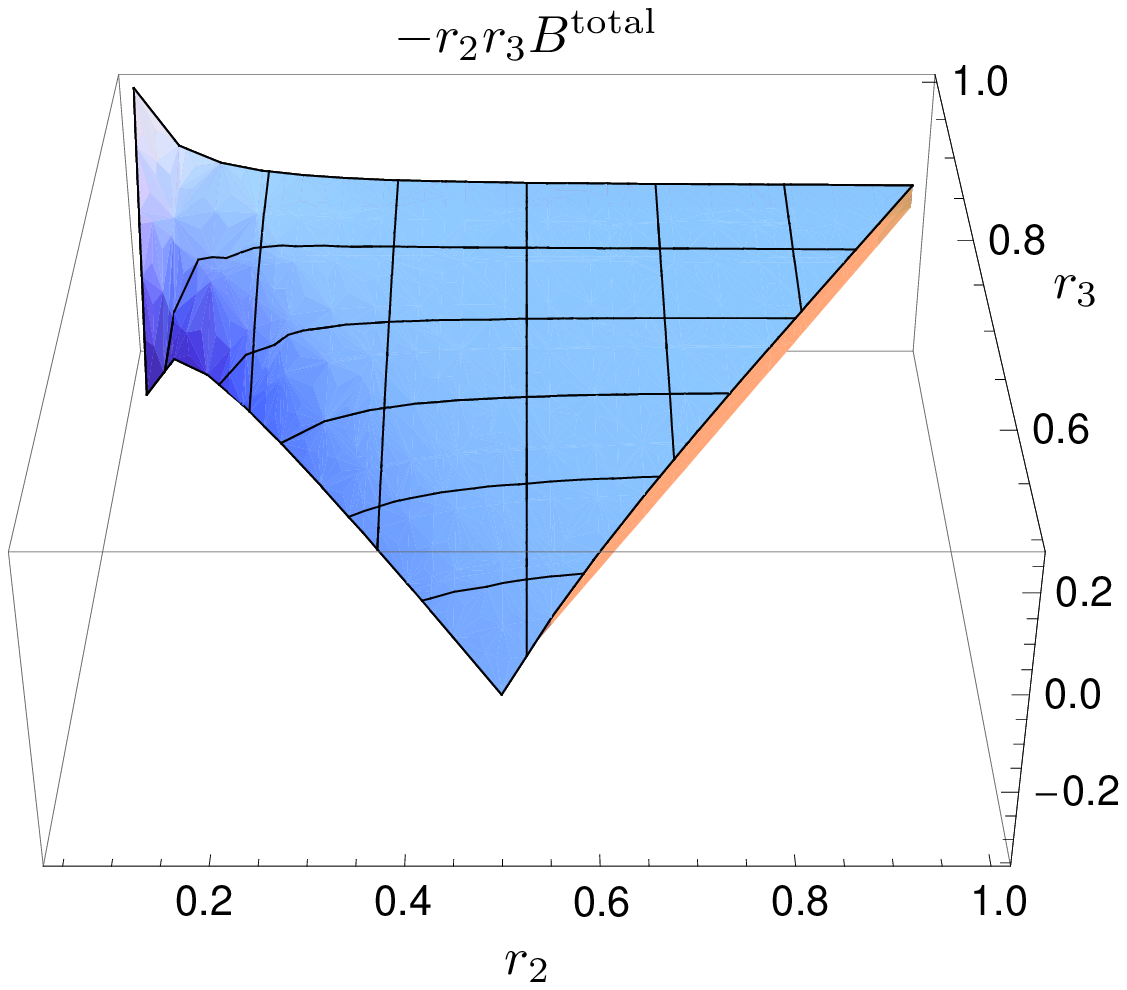}}   &
\raisebox{0.9cm}{\resizebox*{0.48\textwidth}{!}{\includegraphics{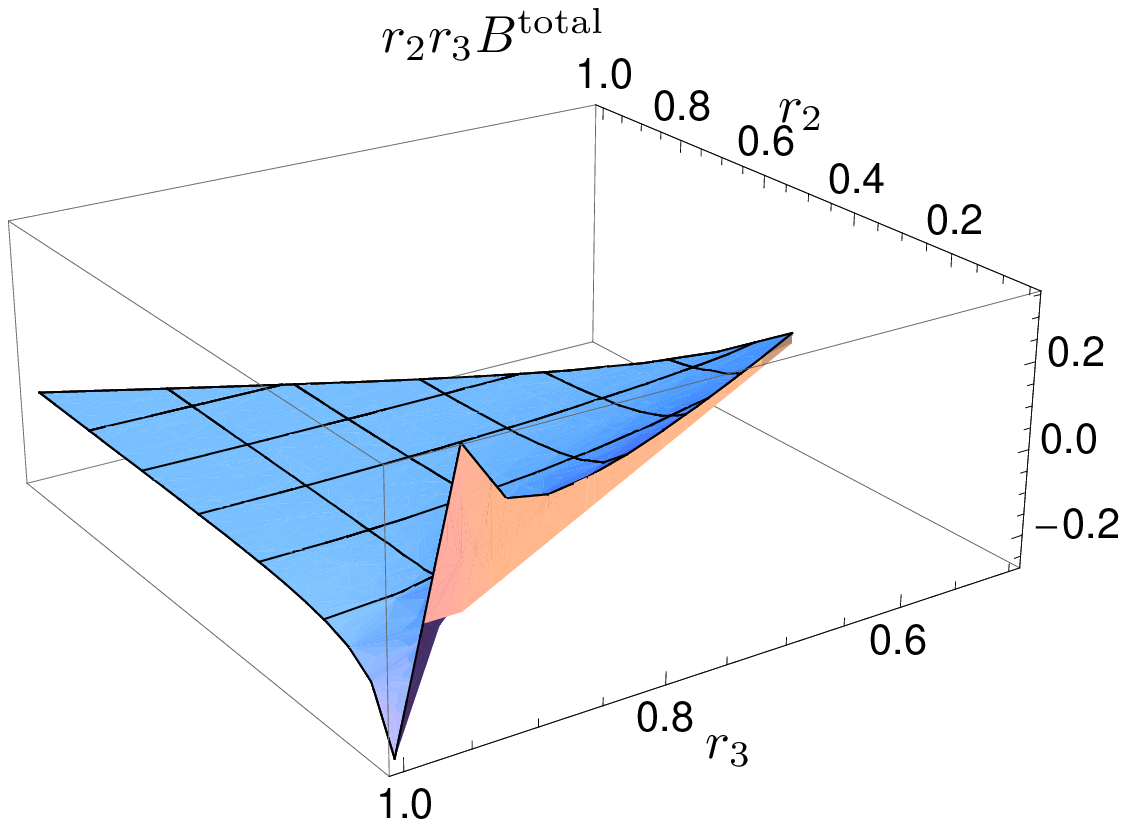}}}
\end{tabular}
\caption{\small The total CMB bispectrum.}
\label{fig:total}
\end{figure}
\end{center}

In the previous section we have separated the calculation of the CMB
bispectrum generated in the Sachs-Wolfe limit into five contributions:
an intrinsic contribution expressed in terms of the Newtonian
potential evaluated at last scattering, in eq.~(\ref{intr_bis}), the
Rees-Sciama effect, in eq.~(\ref{RS}), a contribution from the time
dependence of the vector and tensor components of the metric,
respectively in eqs.~(\ref{V_bis}) and (\ref{T_bis}), and finally the
lensing effect, in eq.~(\ref{eq:lensexplicit}). However, it is
important to stress that only the sum of these contributions has a
physical, gauge invariant, meaning. In this section we turn to discuss this sum, i.e.~the total bispectrum. This is plotted in figure~\ref{fig:total}.
By comparing this with figure \ref{fig:lens} one can appreciate that
the lensing effect largely dominates the total bispectrum \footnote{As
already stressed, the separation among different effects is
gauge-dependent. Here we use the standard terminology in calling
``lensing'' the deflection of photons in Newtonian gauge. For a
discussion about the gauge-dependence of lensing, see \cite{Lewis:2006fu}.}. Let us see this more quantitatively. 

In the squeezed limit the bispectrum is dominated by the intrinsic contribution and the lensing. In this limit we can compare the total bispectrum to the local bispectrum (\ref{eq:local}) taken with $f_{\rm NL}^{\rm local}=1$. This yields
\be
\label{total_loc}
\frac{B^{\rm total}(1,r_2 \to 0,r_3 \to 1)}{B^{\rm local}(1,r_2 \to 0,r_3 \to 1)} = -1/6 - \cos(2 \theta)\;,
\ee
where $\theta$ represents the angle between the short and long wavelength modes $\l_1$ and $\l_2$. Thus, the total bispectrum corresponds to $f_{\rm NL}^{\rm local} = -1/6 - \cos(2 \theta)$.

Note that this result can be obtained by simple arguments. As explained in section~\ref{SWsecord}, the factor $-1/6$ can be inferred using the fact that a mode still out of the Hubble radius today cannot affect a physical measurement. The angular dependent factor $- \cos(2 \theta)$ can be inferred by looking at the effect of a long wavelength lensing mode on the power spectrum, as explained in section~\ref{sec:lensing}.

A remark on the angular dependence in eq.~(\ref{total_loc}) is in
order here. Although it is non-vanishing in the squeezed limit, the
lensing contribution (\ref{eq:lensexplicit}) is not of the local form
(\ref{eq:local}). In particular, as the angular dependence averages to
zero, a non-Gaussianity test based on a local estimator of the form
(\ref{eq:local}) would be almost blind to the lensing signal. A quantitative way to measure how a signal overlaps with another is provided by the cosine between two bispectra, defined as \cite{Babich:2004gb}
\be
\cos(B_1, B_2)  \equiv \frac{B_1 \cdot B_2}{\sqrt{B_1 \cdot B_1 } \sqrt{B_2 \cdot B_2}}\;,
\ee
where $B_1 \cdot B_2$ is the scalar product between two bispectra, given by
\begin{eqnarray}
B_1 \cdot B_2 &\equiv &\frac{1}{\pi} \int \frac{\rmd^2 l_2 \rmd^2 l_3}{(2
  \pi)^2} \frac{B_1(\l_1,\l_2,\l_3) \, B_2(\l_1,\l_2,\l_3)}{6 C_{l_1}
  C_{l_2} C_{l_3}}  \\
& \propto & \int \rmd r_2 \rmd r_3 \frac{r_2^3 r_3^3 B_1(1,r_2,r_3) \,  B_2(1,r_2,r_3)}{\left(2 r_2^2 +2  r_3^2 +2 r_2^2 r_3^2 -1 -r_2^4 -r_3^4 \right)^{1/2}}  \;.
\end{eqnarray}
Indeed, we find that the cosine between the lensing bispectrum
(\ref{eq:lensexplicit}) and the local bispectrum (\ref{eq:local}) is
$\cos(B_{\rm lens}, B_{\rm local}) = 0.03$.\footnote{Note that the
  scalar product with the local bispectrum is logarithmically
  divergent for $r_2 \to 1$ or $r_3 \to 1$. Thus, in order to evaluate
  it we have put the cutoff $r_{\rm max}=0.999$.} For instance, one
can compare this to the cosine between the local and equilateral
bispectra, which is much larger, $\cos(B_{\rm equil}, B_{\rm local}) =
0.30$. Thus, due to the angular dependence of the squeezed limit, the lensing signal is orthogonal to the local one. 
We can now compare the total bispectrum to the local one. The cosine
is $\cos(B_{\rm total}, B_{\rm local}) = - 0.17$. Thus, as it is
dominated by lensing, the total bispectrum is almost orthogonal to the
local signal. However, due to the term $-1/6$ in eq.~(\ref{total_loc})
the orthogonality is not complete and the total bispectrum slightly overlaps with the local one.

In the equilateral limit all the five contributions to the total bispectrum become important. 
However, the lensing numerically dominates. In this limit we can compare the total bispectrum to the equilateral bispectrum (\ref{Bequil}) taken with $f_{\rm NL}^{\rm equil}=1$. This yields
\be
\label{total_eq}
\frac{B^{\rm total}(1,1, 1)}{B^{\rm equil}(1,1, 1)} = 3.13\;.
\ee
Thus, the total bispectrum corresponds to $f_{\rm NL}^{\rm equil} = 3.13$. As it is not vanishing in the squeezed limit, its cosine with the equilateral shape will be smaller than unity. Indeed we find $\cos(B_{\rm total}, B_{\rm equil}) = 0.41$. Note that this value is larger than the cosine between local and equilateral shapes, i.e.~$0.30$. Thus, the total bispectrum is ``more equilateral'' than the local one.  
Finally, to have a confirmation that the lensing effect dominates the
total bispectrum, we can  compute the cosine between the total signal
and the lensing. This is $\cos(B_{\rm total}, B_{\rm lens}) = 0.98$,
which is very close to one, as expected. A summary of the cosines is given in table \ref{table}.
 \begin{table}[t]
 \begin{center}
 \begin{tabular}{|c|c|c|c|c|}
        \hline 
        Shape: & total & local & equil & lens \\
        \hline         \hline 
        total & 1.00 & -0.17 &  0.41 & 0.98 \\
        \hline
	local &   & 1.00 & 0.30 & 0.03 \\
        \hline
	equil &  &  & 1.00 & 0.47 \\
        \hline
	lens &  &  &  & 1.00 \\
        \hline
  \end{tabular}
  \end{center}
  \caption{\label{table} \small Cosines between different shapes of bispectra.}
  \end{table}

It is important to stress that the shape associated with lensing, with
an angle dependent squeezed limit, represents another interesting
template for the bispectrum besides the local, the equilateral and the ones studied in
\cite{Meerburg:2009ys,Senatore:2009gt}. As it is rather orthogonal
to the standard local and equilateral templates, in the future  it would be
interesting to put limits on it, even independently of lensing.

%%%%%%%%%%%%%%%%%%%%%%%%%%%%%%%%%%%%%%%%%%%%%%%
%%%%%%%%%%%%%%%%%%%%%%%%%%%%%%%%%%%%%%%%%%%%%%%
\section{\label{conclusions}Conclusion}
%%%%%%%%%%%%%%%%%%%%%%%%%%%%%%%%%%%%%%%%%%%%%%%
%%%%%%%%%%%%%%%%%%%%%%%%%%%%%%%%%%%%%%%%%%%%%%%

In this paper we have calculated, assuming perfect matter dominance,  the complete CMB bispectrum on large angular
scales, larger than the Hubble radius at recombination, considering
for the first time all the relevant effects. Although our results give the exact
bispectrum in a well defined physical limit, there are many ways to
improve our calculations to make them closer to the real universe. One
should include the recent dark energy domination and the early
transition from radiation to matter dominance along the lines of \cite{Bartolo:2005kv}.
This will give qualitative new phenomena, like the rather large ISW-lensing
correlation \cite{Hanson:2009kg}. Given that we are on large angular scales, a full-sky
treatment would be more precise than our flat-sky expressions,
although the results for the bispectrum will be much more complicated
and difficult to understand. Finally, the small deviation from a scale
invariant spectrum should be included.
Taking all this into account would give the correct prediction for our
universe of the large angle bispectrum. 
This is clearly far from the complete answer. The modes on scales larger than the horizon at
recombination are quite few and most of the bispectrum signal comes
from triangles with modes on sub-Hubble scales. Entering in a
sub-Hubble regime requires the whole machinery of second-order
Boltzmann equations that we have not touched in this paper. 

The calculated bispectrum is rather small:
the final bispectrum is dominated by the lensing contribution, which
gives $f_{\rm NL}^{\rm local} = - \cos (2 \theta)$, with $\theta$ the
angle between long and short modes. Even if we could use our results on
arbitrarily short scales, this would be below Planck sensitivity,
limited to $f_{\rm NL}^{\rm local} \sim 5$. This means that the
bispectrum in the Sachs-Wolfe limit does not represent a relevant
contamination for the forthcoming searches for primordial non-Gaussianities.

A way to go beyond the large angle regime is to correct the results of 
\cite{Creminelli:2004pv} to get the full bispectrum in the squeezed
limit, with one (but not necessarily all) of the modes on scales
larger than the horizon at recombination. We leave all these
directions for future work.

%%%%%%%%%%%%%%%%%%%%%%%%%%%%%%%%%%%%%%%%%%%%%%%
%%%%%%%%%%%%%%%%%%%%%%%%%%%%%%%%%%%%%%%%%%%%%%%
\section*{Acknowledgments}
It is a pleasure to thank Nicola Bartolo, Francis Bernardeau, Eiichiro
Komatsu, Roy Maartens, Sabino Matarrese, Toni Riotto, Misao Sasaki, Uros Seljak, Leonardo Senatore and Matias Zaldarriaga for useful discussions. G.D'A.~and P.C.~thank the Institut de Physique Th\'eorique at Saclay and F.V.~thanks the ICTP for hospitality while working on this project. 
G.D'A., P.C., and F.V.~thank the Galileo Galilei Institute, where part of this work was carried out during the workshop ``New Horizons for Modern Cosmology''.
Furthermore, F.V.~thanks the Yukawa Institute for Theoretical Physics
at Kyoto University, where part of this work was carried out during
the GCOE/YITP workshop YITP-W-09-01 on ``Non-linear cosmological
perturbations'' and the EU Marie Curie Research
and Training network "UniverseNet" (MRTN-CT-2006-035863) for support..

\appendix

\section*{Appendix}
\section{Flat-sky and integrated effects}
\label{app:flat_sky}

At first order, the gravitational contribution to the temperature
anisotropies in matter domination is the Sachs-Wolfe effect,
\begin{equation}
  \frac{\del T}{T} (\hat n) = \frac{1}{3} \phi(\n D_e)
  = \frac{1}{3} \int \frac{\rmd^3 k}{(2 \pi)^3} e^{i \k \cdot \n D_e}
  \phi_{\k} \, ,
\end{equation}
where $\n$ is the unit vector specifying the line of sight direction, $D_e = \tau_o-\tau_e$ is the (background) conformal distance
to the last scattering surface and $\phi$ is the first order
Newtonian potential.
In the flat-sky formalism~\cite{Seljak:1995ve, Hu:2000ee}, one chooses a
fiducial direction $\hat{z}$ and expands at the lowest
order in the angle $\th$ between $\hat{z}$ and $\n$:
\begin{equation}
  \n = (\sin \th \cos \phi, \sin \th \sin \phi, \cos \th)
  \simeq (m_x, m_y, 1) \, ,
\end{equation}
$\vec{m}$ being a 2-dimensional vector normal to $\hat{z}$.
The multipole is simply the 2-dimensional Fourier transform with
respect to $\vec{m}$:
\begin{equation}
\label{SW}
  a_{\l} = \int \rmd^2 m\; e^{- i \l \cdot \vec{m}} \frac{\del T}{T}(\n)
  = \frac{1}{3} \int \frac{\rmd^3 k}{2 \pi} \delta(\l - \k_\parallel D_e)
  e^{i k_\perp D_e} \phi_{\k} \, .
\end{equation}
One can show that the flat-sky multipole corresponds to the large $l$ limit of the full-sky one. The two are related by~\cite{Hu:2000ee}
\begin{align}
a_{\l} = \sqrt{\frac{4 \pi}{2 l +1}} \sum_m i^{-m} a_{l m} e^{i m \varphi_l} \;,\\
a_{l m} = \sqrt{\frac{2 l +1}{4 \pi}} i^m \int \frac{\rmd \varphi_l}{2
  \pi} 
e^{- i m \varphi_l} a_{\l} \;.
\end{align}
Similar expressions hold also for the power spectrum and the bispectrum.
The power spectrum is defined as
$\langle a_{\l_1} a_{\l_2} \rangle \equiv (2\pi)^2\delta(\l_1+\l_2) C_{l_1}^{\rm
flat}$ in flat-sky approximation, and as
$\langle a_{l_1 m_1} a_{l_2 m_2} \rangle \equiv
\delta_{m_1 m_2} \delta_{l_1 l_2} C_{l_1}^{\rm full}$ in full sky;
the two expressions are related by $C_l^{\rm full} \approx C_l^{\rm flat}$ for large $l$.
The bispectrum in the full and flat sky are defined respectively as
\be
\langle a_{\l_1} a_{\l_2} a_{\l_3} \rangle \equiv
(2\pi)^2 \delta(\l_1 + \l_2 + \l_3) B(\l_1,\l_2,\l_3) \; ,
\ee
\be
\ens{a_{l_1 m_1} a_{l_2 m_2} a_{l_3 m_3}}
\equiv  \left( \begin{matrix} l_1 & l_2 & l_3 \\ m_1 &
  m_2 & m_3 \end{matrix} \right) B_{l_1 l_2 l_3} \; ,
\ee
where $\bigl( \begin{smallmatrix} l_1 & l_2 & l_3 \\ m_1 &
  m_2 & m_3 \end{smallmatrix} \bigr)$ is the Wigner 3-j symbol.
The two expressions are related by:
\begin{equation}
B_{l_1 l_2 l_3} \approx \begin{pmatrix}
l_1 & l_2 & l_3 \\ 0 &
  0 & 0 \end{pmatrix} \sqrt{\frac{(2l_1 + 1)(2l_2 + 1)(2l_3 +
    1)}{4\pi}} B(\l_1,\l_2,\l_3) \, .
\end{equation}
The derivation of these expressions can be found in~\cite{Hu:2000ee}.

To better understand what happens when we correlate effects which are
important at different times, we can do a simple exercise\footnote{We
  thank F.~Bernardeau for suggesting this example.}: we calculate the
2-point function of two integrated effects which peak at different
times $\tau_1$ and $\tau_2$.
We will see that the correlation decays exponentially when
$\tau \gtrsim (\tau_2-\tau_1)/l$, and that the power spectrum is
proportional to $\delta(\l_1+\l_2)$ up to exponentially small terms.
Consider a generic integrated effect at first order:
\begin{equation}
  a_{\l} = \int_{\tau_e}^{\tau_o} \rmd \tau \int \frac{\rmd^3 k}{2 \pi}
  \delta(\l - \k_\parallel (\tau_o - \tau))
  e^{i k_\perp (\tau_o - \tau)} g'(\tau) \phi_{\k} \, ,
\end{equation}
where $g(\tau)$ is a growth function.
Now we correlate two such effects, with different growth functions $g(\tau)$ and $f(\tau)$:
\begin{equation}
\begin{split}
  \langle a^f_{\l_1} a^g_{\l_2}\rangle
  &= 4 \pi \int_{\tau_e}^{\tau_o} \rmd \tau_a \, f'(\tau_a)
  \int_{\tau_e}^{\tau_o} \rmd \tau_b \, g'(\tau_b)
  \int \rmd^3 k \delta(\l_1 - \k_\parallel (\tau_o-\tau_a)) \delta(\l_2 + \k_\parallel (\tau_o - \tau_b))
  e^{i k_\perp (\tau_b- \tau_a)} \frac{A}{k^3} \\
  &= 4 \pi \int_{\tau_e}^{\tau_o} \rmd \tau_a \,\frac{f'(\tau_a)}{(\tau_o-\tau_a)^2}
  \int_{\tau_e}^{\tau_o} \rmd \tau_b \, g'(\tau_b)
  \int \rmd k_\perp
  \delta\left(\l_1 + \l_2 + \l_1 \frac{\tau_a - \tau_b}{\tau_o -
      \tau_a}\right) \\
  &\phantom{=}\times e^{i k_\perp (\tau_b - \tau_a)}
  \frac{A}{(k_\perp^2+l_1^2/(\tau_o-\tau_a)^2)^{3/2}} \;.
\end{split}
\end{equation}
For simplicity, we approximate the growth functions with step functions, such that
\begin{equation}
  f'(\tau) \sim \del(\tau - \tau_1) \, ,
  \qquad g'(\tau) \sim \del(\tau - \tau_2) \;,
\end{equation}
where we consider $\tau_e \leq \tau_1 \leq \tau_2 < \tau_o$.
Thus we find
\begin{equation}
  \langle a^f_{\l_1} a^g_{\l_2}\rangle
  = \frac{4 \pi A}{(\tau_o-\tau_1)^2}
   \delta\left(\l_1 + \l_2 + \l_1 \frac{\tau_1 - \tau_2}{\tau_o -
      \tau_1}\right)  
  \int \rmd k_\perp e^{i k_\perp (\tau_2 - \tau_1)}
  (k_\perp^2 + l_1^2/(\tau_o-\tau_1)^2)^{-3/2} \, .
\end{equation}
The integration over $k_\perp$ can be done analytically, yielding
\begin{equation}
  \langle a^f_{\l_1} a^g_{\l_2}\rangle 
  = (2 \pi)^2 \delta\left(\l_1 + \l_2 + \l_1 \frac{\tau_1 - \tau_2}{\tau_o -
      \tau_1}\right)
  \frac{2}{\pi} \frac{|\tau_2-\tau_1|}{(\tau_o-\tau_1)}
  \frac{A}{l_1} K_1 \left(l_1 \frac{|\tau_2-\tau_1|}{\tau_o-\tau_1}\right) \, ,
\end{equation}
where $K_1$ is the modified Bessel function, with asymptotic behaviours
$K_1(x) \to 1/x$ for $x \ll \sqrt{2}$ and $K_1(x) \to \sqrt{\pi/ 2 x} \, e^{-x}$
for $x \gg 3/4$.
\begin{figure}[htp]
\centering
\includegraphics{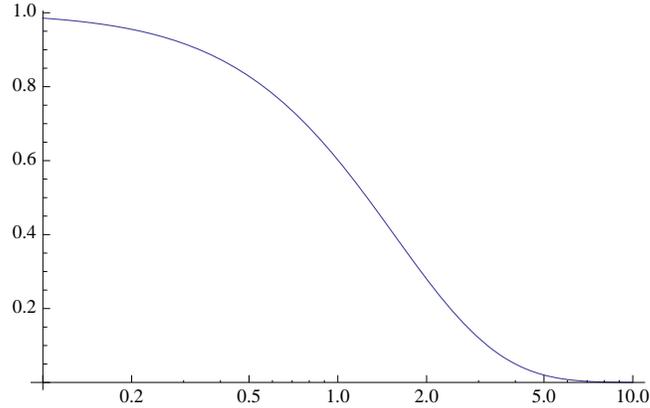}
\caption{\small{The function $x K_1(x)$ with the $x$ axis in logarithmic scale.}}\label{fig:K1}
\end{figure}

We can see that,
when $l_1 \frac{\tau_2-\tau_1}{\tau_o-\tau_1} \gtrsim 1$,
the correlation is exponentially suppressed.
Then, in the limit $l_1 \ll (\tau_o-\tau_1)/(\tau_2-\tau_1)$
we find
\begin{equation}
  \langle a^f_{\l_1} a^g_{\l_2}\rangle 
  = (2 \pi)^2 \delta(\l_1 + \l_2)
  \frac{2}{\pi} \frac{A}{l_1^2} \,.
\end{equation}
In general, the translational invariance in 2d is only approximate;
however, the approximation is very good since for large multipoles the correlations are exponentially suppressed if the sum $\sum_i \l_i \neq 0$.

\section{Detailed calculation of the Rees-Sciama effect}
\label{app:RS}

In this appendix we compute the Rees-Sciama bispectrum.
We start from eq.~\eqref{RS} before summing over cyclic permutations:
\begin{equation}
  \begin{split}
    B^{\mathrm{RS}}
%    = - \frac{A^2}{378 \pi^2} \frac{1}{l_1^4} \int_{l_1 \tau_e/D_e}^{l_1 \tau_o/D_e} \rmd u \, u
%    \int_{-\infty}^{+\infty} \rmd y_1 \int_{-\infty}^{+\infty} \rmd y_2
%    e^{i (y_1 + y_2) (u - u_r)} (y_1^2 + r_1^2)^{-3/2}
%    (y_2^2 + r_2^2)^{-3/2} \times \\
%    \times \left[ \frac{5}{2} (y_1^2 + y_2^2) + 2 y_1 y_2 
%      + \frac{3}{2} r_1^2 + \frac{3}{2} r_2^2 + r_3^2
%      - \frac{5}{2}
%      \frac{(y_1^2 - y_2^2 + r_1^2 - r_2^2)^2}{(y_1+y_2)^2 + r_3^2} \right] = \\
    = - \frac{A^2}{378 \pi^2} \frac{1}{l_1^4}
    \int_0^{\infty} \rmd x \, x
    \int_{-\infty}^{+\infty} \rmd y_1 \rmd y_2\,
    e^{i (y_1 + y_2) x} (y_1^2 + r_1^2)^{-3/2}
    (y_2^2 + r_2^2)^{-3/2} \\
    \times \left[\frac{3}{2} r_1^2 + \frac{3}{2} r_2^2 + r_3^2
      + 2 y_1 y_2 + \frac{5}{2} (y_1^2 + y_2^2)
      - \frac{5}{2}
      \frac{(y_1^2 - y_2^2 + r_1^2 - r_2^2)^2}{(y_1+y_2)^2 + r_3^2}
    \right] \, .
\label{eq:RSapp}
  \end{split}
\end{equation}
Since there are some pieces in the kernel that can be integrated
analytically, we now compute them as a check of our numerical integration.
To proceed, we make use of the following known integrals:
\begin{align}
  & \int_{-\infty}^{+\infty} \rmd y \frac{e^{i y x}}{(y^2 + a^2)^{\frac{3}{2}}}
  = \frac{2 x}{a} K_1(a x) \, ,\\
  & \int_{-\infty}^{+\infty} \rmd y e^{i y x} \frac{y}{(y^2 + a^2)^{\frac{3}{2}}} = - i \frac{\rmd}{\rmd x} \left( \frac{2 x}{a} K_1(a x) \right)
  = 2 i x K_0(a x) \, ,\\
  & \int_{-\infty}^{+\infty} \rmd y e^{i y x} \frac{y^2}{(y^2 + a^2)^{\frac{3}{2}}} = - i \frac{\rmd}{\rmd x} \left(2 i x K_0(a x) \right)
  = 2 \left[K_0(a x) - a x K_1(a x) \right] \, ,
\end{align}
where the $K_i$ are Bessel modified functions. We then split the bispectrum into four pieces, three of which are integrated analytically:
\begin{equation}
  B^{(1)} = - \frac{2 A^2}{189 \pi^2} \frac{1}{l_1^4}
  \left( \frac{3}{2} r_1^2 + \frac{3}{2} r_2^2 + r_3^2 \right)
  \frac{1}{r_1 r_2}
  \int_0^{\infty} \rmd x \, x^3 K_1(r_1 x) K_1(r_2 x)\,,
\end{equation}
\begin{equation}
  B^{(2)} =  \frac{4 A^2}{189 \pi^2} \frac{1}{l_1^4}
  \int_0^{\infty} \rmd x \, x^3 K_0(r_1 x) K_0(r_2 x)\,,
\end{equation}
\begin{multline}
    B^{(3)} 
%    = 
%    - \frac{5 A^2}{189 \pi^2} \frac{1}{l_1^4}
%     \int_0^{\infty} \rmd x \, x
%    \times \bigg[ \frac{x}{r_1} K_1(r_1 x) \left(K_0(r_2 x) - r_2 x
%        K_1(r_2 x) \right) \\
%      +& \left(K_0(r_1 x) - r_1 x K_1(r_1 x) \right) \frac{x}{r_2} K_1(r_2 x) \bigg] \\
    = - \frac{5 A^2}{189 \pi^2} \frac{1}{l_1^4}
    \int_0^{\infty} \rmd x \, x \Bigg[ x \left(
        \frac{1}{r_1} K_1(r_1 x) K_0(r_2 x) + \frac{1}{r_2} K_0(r_1 x)
        K_1(r_2 x)\right) \\
      - x^2 \left( \frac{r_2}{r_1} +
        \frac{r_1}{r_2}\right) K_1(r_1 x) K_1(r_2 x) \Bigg]\,,
\end{multline}
\begin{equation}
  B^{(4)} =  \frac{5 A^2}{756 \pi^2} \frac{1}{l_1^4}
  \int_0^{\infty} \rmd x \, x
  \int_{-\infty}^{+\infty} \rmd y_1 \rmd y_2\,
  e^{i (y_1+y_2) x}
  (y_1^2 + r_1^2)^{-\frac{3}{2}} (y_2^2 + r_2^2)^{-\frac{3}{2}}
  \frac{(y_1^2 - y_2^2 + r_1^2 - r_2^2)^2}{(y_1+y_2)^2 + r_3^2}\,.
  \label{B4}
\end{equation}
%
%First of all, we push the upper limit of integration to $+ \infty$
%since the integrand is exponentially suppressed when $x > 1$. 
%Another important simplification consists in considering $l_1 \tau_e/D_e \ll 1$,
%which means $k_{1 \parallel} \gg \tau_e^{-1}$, that is the mode should be out
%of the Hubble radius at recombination.  Then, since the integrals are peaked
%around $x \sim 1$, we can approximate in the integrand functions $(l_1
%\tau_e/D_e + x) \simeq x$.
%With these approximations, we find
%
The first three pieces can be integrated in $x$, giving:
\begin{equation}
  B^{(1)} = - \frac{4 A^2}{189 \pi^2} \frac{1}{l_1^4}
  \left( \frac{3}{2} r_1^2 + \frac{3}{2} r_2^2 + r_3^2 \right)
  \frac{1}{r_1^2 r_2^2 (r_2^2-r_1^2)^3} \left[r_2^4 - r_1^4 
  - 4 r_1^2 r_2^2 \ln{\frac{r_2}{r_1}} \right]\, ,
  \label{B1}
\end{equation}
\begin{equation}
  B^{(2)} =  \frac{16 A^2}{189 \pi^2} \frac{1}{l_1^4}
\frac{1}{(r_2^2-r_1^2)^3} \left[r_1^2 - r_2^2 
  - (r_1^2 + r_2^2) \ln{\frac{r_1}{r_2}} \right]\, ,
\end{equation}
\begin{equation}
    B^{(3)} = - \frac{5 A^2}{189 \pi^2} \frac{1}{l_1^4}
    \frac{1}{(r_1^2-r_2^2)^3} \bigg[ 5 (r_2^2 - r_1^2) +
        \frac{r_2^4}{r_1^2} - \frac{r_1^4}{r_2^2} + 2 (r_1^2 - r_2^2)
        \ln{\frac{r_1}{r_2}} - 2 (3 r_1^2 + 5 r_2^2)
        \ln{\frac{r_2}{r_1}} \bigg]\, .
  \label{B3}
\end{equation}
%
%\begin{multline}
%  B^{(4)}(\l_1,\l_2,\l_3) =  \frac{5 A^2}{756 \pi^2} \frac{1}{l_1^4}
%  \int_0^{\infty} \rmd x \, x
%  \int_{-\infty}^{+\infty} \rmd y_1 \int_{-\infty}^{+\infty} \rmd y_2\,
%  e^{i (y_1+y_2) x} \\
%  \times (y_1^2 + r_1^2)^{-\frac{3}{2}} (y_2^2 + r_2^2)^{-\frac{3}{2}}
%  \frac{(y_1^2 - y_2^2 + r_1^2 - r_2^2)^2}{(y_1+y_2)^2 + r_3^2}\, .
%  \label{B4}
%\end{multline}
%
The fourth piece \eqref{B4} cannot be integrated analytically, making the numerical integration necessary. However, comparison between the numerical integration of the other three pieces and the analytical expressions \eqref{B1} to \eqref{B3} gives consistent results. This provides a check of the validity of our computation.

Now we turn to the numerical integration of eq.~\eqref{eq:RSapp}. To do it we first have to perform analytically
the $x$ integral, which is ill-defined. In order to overcome this
problem we first change variables from $y_1$, $y_2$ to $y_{+} \equiv
y_1 + y_2$ and $y_{-} \equiv y_1 - y_2$, and then regularize the
integral in the following way:
\begin{align}
  \int_0^{\infty} \rmd x \int _{-\infty}^{+\infty} \rmd y_+
  \int_{-\infty}^{+\infty}\rmd y_- \,x e^{i y_+ x} f(y_+,y_-)
  &= - i \int _{-\infty}^{+\infty} \rmd y_+
  \int _{-\infty}^{+\infty} \rmd y_- \int_0^{\infty} \rmd x
  f(y_+,y_-) \frac{\de}{\de y_+} e^{i y_+ x} \nonumber \\
  &= i \int _{-\infty}^{+\infty} \rmd y_+
  \int _{-\infty}^{+\infty} \rmd y_- \frac{\de}{\de y_+} f(y_+,y_-)
  \int_0^{\infty} \rmd x\,   e^{i y_+ x} \nonumber \\
  &= - \int _{-\infty}^{+\infty} \rmd y_+
  \int _{-\infty}^{+\infty} \rmd y_- \frac{1}{y_+} \frac{\de}{\de y_+} f(y_+,y_-) \, .
  \label{eq:xint}
\end{align}
The last integral follows from the prescription
\begin{equation}
  \int_0^{\infty} \rmd x   e^{i y_+ x} e^{-\eps x} = \frac{1}{-i y_+ + \eps} \, .
\end{equation}
After integrating in $x$, we obtain:
\begin{multline}
    B^{\mathrm{RS}}
    = \frac{A^2}{378 \pi^2} \frac{1}{l_1^4}
    \int_{-\infty}^{+\infty} \rmd y_1 \rmd y_2\,
    \frac{1}{y_+}\frac{\partial}{\partial y_+} \Bigg\{(y_1^2 + r_1^2)^{-3/2}
    (y_2^2 + r_2^2)^{-3/2} \\
    \times \left[\frac{3}{2} r_1^2 + \frac{3}{2} r_2^2 + r_3^2
      + 2 y_1 y_2 + \frac{5}{2} (y_1^2 + y_2^2)
      - \frac{5}{2}
      \frac{(y_1^2 - y_2^2 + r_1^2 - r_2^2)^2}{(y_1+y_2)^2 + r_3^2}
    \right]\Bigg\} \, ,
\end{multline}
which, after changing variables from $(y_1,y_2)$ to $(y_+,y_-)$, and
performing the derivative, gives a form which can be integrated
numerically. The final results of the integration, after summing over cyclic permutations and setting $r_1=1$, are presented in
figure \ref{fig:RS}.

%{\bf Local shape and squeezed limit.} To study the 3-point
%function induced by the Rees-Sciama effect, it is useful to compare it with
%the one induced by a primordial $f_{\rm NL}^{\rm local}$.

%The RS gives for an equilateral configuration
%\begin{equation}
%- 0.012 \cdot \frac{A^2}{l_1^4}  \;.
%\end{equation}
%While $f_{\rm NL}^{\rm local}$ gives
%\begin{equation}
%- \frac{2}{9 \pi^2} \frac{A^2}{l_1^4} \cdot  f_{\rm NL}^{\rm
%  local} \simeq - 0.023 \cdot \frac{A^2}{l_1^4} \cdot  f_{\rm NL}^{\rm
%  local}\;.
%\end{equation}
%Thus the Rees-Sciama contribution corresponds to $f_{\rm NL}^{\rm
%  local} \simeq 1/2$ in the equilateral configuration.

We can compare the Rees-Sciama contribution with the local shape in
the squeezed limit.  We will see that while the local shape diverges
as $1/r^2$ in this limit, the Rees-Sciama only diverges as $1/r$.
Going back to eq.~(\ref{eq:RSapp}), we can study the behavior when one
of the $r$ goes to zero; notice that the expression must be
symmetrized so that we have to study both the limits $r_2 \to 0$ and
$r_3 \to 0$ in eq.~(\ref{eq:RSapp}). For $r_2 \to 0$ we have an
infrared divergence in the $y_2$ integral coming from the power
spectrum which goes as $y_2^{-3}$ for $r_2=0$. This would give a
divergence $r_2^{-2}$ as in the local model. However, for $r_2=0$ and
$r_1 = r_3 =1$ the expression in brackets in the second line of
(\ref{eq:RSapp}) goes as $y_2$ for $y_2 \to 0$, but its integral
vanishes due to parity, leaving only terms which are at most
logarithmic divergent and thus suppressed with respect to the local
shape. An additional divergence comes from the limit $r_3 \to 0$ in
eq.~(\ref{eq:RSapp}); in this case the integral diverges in the limit
$y_1 +y_2 \to 0$. Notice that in this case one also has to take into
account the integral over $x$ which diverges for $y_1+y_2 = 0$. To
study the behavior for $r_3 \to 0$ one must first integrate in $x$
using the prescription \eqref{eq:xint}. One can see that the leading
divergence in the resulting expression comes from a term of the form
$r_3^2/y_+^4$, which gives a $1/r_3$ divergence. This is dominant
compared to the divergence in $r_2$, but it is still subdominant
compared to the local case. We conclude that the Rees-Sciama result is
subdominant compared to the local shape in the squeezed limit. This
analysis is a good check of the numerics, which indeed shows a $1/r$
divergence in the squeezed limit.


\footnotesize
\parskip 0pt

\begin{thebibliography}{99}

%\cite{Komatsu:2008hk}
\bibitem{Komatsu:2008hk}
  E.~Komatsu {\it et al.}  [WMAP Collaboration],
  ``Five-Year Wilkinson Microwave Anisotropy Probe
  Observations:Cosmological Interpretation,''
  Astrophys.\ J.\ Suppl.\  {\bf 180}, 330 (2009)
  [arXiv:0803.0547 [astro-ph]].
  %%CITATION = APJSA,180,330;%%

%\cite{Slosar:2008hx}
\bibitem{Slosar:2008hx}
  A.~Slosar, C.~Hirata, U.~Seljak, S.~Ho and N.~Padmanabhan,
  ``Constraints on local primordial non-Gaussianity from large scale
  structure,''
  JCAP {\bf 0808}, 031 (2008)
  [arXiv:0805.3580 [astro-ph]].
  %%CITATION = JCAPA,0808,031;%%

%\cite{Smith:2009jr}
\bibitem{Smith:2009jr}
  K.~M.~Smith, L.~Senatore and M.~Zaldarriaga,
  ``Optimal limits on $f_{NL}^{local}$ from WMAP 5-year data,''
  arXiv:0901.2572 [astro-ph].
  %%CITATION = ARXIV:0901.2572;%%
  
  %\cite{Seljak:1998nu}
\bibitem{Seljak:1998nu}
  U.~Seljak and M.~Zaldarriaga,
  ``Direct Signature of Evolving Gravitational Potential from Cosmic Microwave
  Background,''
  Phys.\ Rev.\  D {\bf 60}, 043504 (1999)
  [arXiv:astro-ph/9811123].
  %%CITATION = PHRVA,D60,043504;%%
  
  %\cite{Goldberg:1999xm}
\bibitem{Goldberg:1999xm}
  D.~M.~Goldberg and D.~N.~Spergel,
  ``Microwave background bispectrum. 2. A probe of the low redshift universe,''
  Phys.\ Rev.\  D {\bf 59}, 103002 (1999)
  [arXiv:astro-ph/9811251].
  %%CITATION = PHRVA,D59,103002;%%
 
 %\cite{Khatri:2008kb}
\bibitem{Khatri:2008kb}
  R.~Khatri and B.~D.~Wandelt,
  ``Crinkles in the last scattering surface: Non-Gaussianity from inhomogeneous
  recombination,''
  Phys.\ Rev.\  D {\bf 79}, 023501 (2009)
  [arXiv:0810.4370 [astro-ph]].
  %%CITATION = PHRVA,D79,023501;%%

%\cite{Senatore:2008vi}
\bibitem{Senatore:2008vi}
  L.~Senatore, S.~Tassev and M.~Zaldarriaga,
  ``Cosmological Perturbations at Second Order and Recombination Perturbed,''
  arXiv:0812.3652 [astro-ph].
  %%CITATION = ARXIV:0812.3652;%%
 
 %\cite{Senatore:2008wk}
\bibitem{Senatore:2008wk}
  L.~Senatore, S.~Tassev and M.~Zaldarriaga,
  ``Non-Gaussianities from Perturbing Recombination,''
  arXiv:0812.3658 [astro-ph].
  %%CITATION = ARXIV:0812.3658;%%
  
%\cite{Pitrou:2008ak}
\bibitem{Pitrou:2008ak}
  C.~Pitrou, J.~P.~Uzan and F.~Bernardeau,
  ``Cosmic microwave background bispectrum on small angular scales,''
  Phys.\ Rev.\  D {\bf 78}, 063526 (2008)
  [arXiv:0807.0341 [astro-ph]].
  %%CITATION = PHRVA,D78,063526;%%

%\cite{Bartolo:2008sg}
\bibitem{Bartolo:2008sg}
  N.~Bartolo and A.~Riotto,
  ``On the non-Gaussianity from Recombination,''
  JCAP {\bf 0903}, 017 (2009)
  [arXiv:0811.4584 [astro-ph]].
  %%CITATION = JCAPA,0903,017;%%

%\cite{Bartolo:2006fj}
\bibitem{Bartolo:2006fj}
  N.~Bartolo, S.~Matarrese and A.~Riotto,
  ``CMB Anisotropies at Second-Order II: Analytical Approach,''
  JCAP {\bf 0701}, 019 (2007)
  [arXiv:astro-ph/0610110].
  %%CITATION = JCAPA,0701,019;%%

%\cite{Nitta:2009jp}
\bibitem{Nitta:2009jp}
  D.~Nitta, E.~Komatsu, N.~Bartolo, S.~Matarrese and A.~Riotto,
  ``CMB anisotropies at second order III: bispectrum from products of the
  first-order perturbations,''
  arXiv:0903.0894 [astro-ph.CO].
  %%CITATION = ARXIV:0903.0894;%%

%\cite{Pitrou:2008hy}
\bibitem{Pitrou:2008hy}
  C.~Pitrou,
  ``The radiative transfer at second order: a full treatment of the Boltzmann
  equation with polarization,''
  Class.\ Quant.\ Grav.\  {\bf 26}, 065006 (2009)
  [arXiv:0809.3036 [gr-qc]].
  %%CITATION = CQGRD,26,065006;%%

%\cite{Bartolo:2004ty}
\bibitem{Bartolo:2004ty}
  N.~Bartolo, S.~Matarrese and A.~Riotto,
  ``Gauge-invariant temperature anisotropies and primordial non-Gaussianity,''
  Phys.\ Rev.\ Lett.\  {\bf 93}, 231301 (2004)
  [arXiv:astro-ph/0407505].
  %%CITATION = PRLTA,93,231301;%%

%\cite{Bartolo:2005fp}
\bibitem{Bartolo:2005fp}
  N.~Bartolo, S.~Matarrese and A.~Riotto,
  ``Non-Gaussianity of Large-Scale CMB Anisotropies beyond Perturbation
  Theory,''
  JCAP {\bf 0508}, 010 (2005)
  [arXiv:astro-ph/0506410].
  %%CITATION = JCAPA,0508,010;%%

%\cite{Bartolo:2005kv}
\bibitem{Bartolo:2005kv}
  N.~Bartolo, S.~Matarrese and A.~Riotto,
  ``The Full Second-Order Radiation Transfer Function for Large-Scale CMB
  Anisotropies,''
  JCAP {\bf 0605}, 010 (2006)
  [arXiv:astro-ph/0512481].
  %%CITATION = JCAPA,0605,010;%%

%\cite{Sachs:1967er}
\bibitem{Sachs:1967er}
  R.~K.~Sachs and A.~M.~Wolfe,
  ``Perturbations of a cosmological model and angular variations of the
  microwave background,''
  Astrophys.\ J.\  {\bf 147}, 73 (1967).
  %%CITATION = ASJOA,147,73;%%

%\cite{Matarrese:1997ay}
\bibitem{Matarrese:1997ay}
  S.~Matarrese, S.~Mollerach and M.~Bruni,
  ``Second-order perturbations of the Einstein-de Sitter universe,''
  Phys.\ Rev.\  D {\bf 58}, 043504 (1998)
  [arXiv:astro-ph/9707278].
  %%CITATION = PHRVA,D58,043504;%%

%\cite{Boubekeur:2008kn}
\bibitem{Boubekeur:2008kn}
  L.~Boubekeur, P.~Creminelli, J.~Nore\~na and F.~Vernizzi,
  ``Action approach to cosmological perturbations: the 2nd order metric in
  matter dominance,''
  JCAP {\bf 0808} (2008) 028
  [arXiv:0806.1016 [astro-ph]].
  %%CITATION = JCAPA,0808,028;%%

%\cite{Mollerach:1995sw}
\bibitem{Mollerach:1995sw}
  S.~Mollerach, A.~Gangui, F.~Lucchin and S.~Matarrese,
  ``Contribution to the three point function of the cosmic microwave background
  from the Rees-Sciama effect,''
  Astrophys.\ J.\  {\bf 453}, 1 (1995)
  [arXiv:astro-ph/9503115].
  %%CITATION = ASJOA,453,1;%%

%\cite{Munshi:1995eh}
\bibitem{Munshi:1995eh}
  D.~Munshi, T.~Souradeep and A.~A.~Starobinsky,
  ``Skewness Of Cosmic Microwave Background Temperature Fluctuations Due To
  Nonlinear Gravitational Instability,''
  Astrophys.\ J.\  {\bf 454}, 552 (1995)
  [arXiv:astro-ph/9501100].
  %%CITATION = ASJOA,454,552;%%

%\cite{Creminelli:2004pv}
\bibitem{Creminelli:2004pv}
  P.~Creminelli and M.~Zaldarriaga,
  ``CMB 3-point functions generated by non-linearities at recombination,''
  Phys.\ Rev.\  D {\bf 70}, 083532 (2004)
  [arXiv:astro-ph/0405428].
  %%CITATION = PHRVA,D70,083532;%%

%\cite{Babich:2004gb}
\bibitem{Babich:2004gb}
  D.~Babich, P.~Creminelli and M.~Zaldarriaga,
  ``The shape of non-Gaussianities,''
  JCAP {\bf 0408}, 009 (2004)
  [arXiv:astro-ph/0405356].
  %%CITATION = JCAPA,0408,009;%%

%\cite{Pyne:1995bs}
\bibitem{Pyne:1995bs}
  T.~Pyne and S.~M.~Carroll,
  ``Higher-Order Gravitational Perturbations of the Cosmic Microwave
  Background,''
  Phys.\ Rev.\  D {\bf 53}, 2920 (1996)
  [arXiv:astro-ph/9510041].
  %%CITATION = PHRVA,D53,2920;%%

%\cite{Mollerach:1997up}
\bibitem{Mollerach:1997up}
  S.~Mollerach and S.~Matarrese,
  ``Cosmic microwave background anisotropies from second order  gravitational
  perturbations,''
  Phys.\ Rev.\  D {\bf 56}, 4494 (1997)
  [arXiv:astro-ph/9702234].
  %%CITATION = PHRVA,D56,4494;%%

%\cite{Misner:1974qy}
\bibitem{Misner:1974qy}
  For a pedagogical introduction to Liouville's theorem in General
  Relativity see C.~W.~Misner, K.~S.~Thorne and J.~A.~Wheeler,
  ``Gravitation,''
%\href{http://www.slac.stanford.edu/spires/find/hep/www?irn=6627595}{SPIRES entry}
{\it  San Francisco 1973, 1279p}.

%\cite{Maldacena:2002vr}
\bibitem{Maldacena:2002vr}
  J.~M.~Maldacena,
  ``Non-Gaussian features of primordial fluctuations in single field
  inflationary models,''
  JHEP {\bf 0305}, 013 (2003)
  [arXiv:astro-ph/0210603].
  %%CITATION = JHEPA,0305,013;%%

%\cite{Acquaviva:2002ud}
\bibitem{Acquaviva:2002ud}
  V.~Acquaviva, N.~Bartolo, S.~Matarrese and A.~Riotto,
  ``Second-order cosmological perturbations from inflation,''
  Nucl.\ Phys.\  B {\bf 667}, 119 (2003)
  [arXiv:astro-ph/0209156].
  %%CITATION = NUPHA,B667,119;%%

%\cite{Langlois:2006vv}
\bibitem{Langlois:2006vv}
  D.~Langlois and F.~Vernizzi,
  ``Nonlinear perturbations of cosmological scalar fields,''
  JCAP {\bf 0702}, 017 (2007)
  [arXiv:astro-ph/0610064].
  %%CITATION = JCAPA,0702,017;%%

%\cite{Hu:2001yq}
\bibitem{Hu:2001yq}
  W.~Hu and A.~Cooray,
  ``Gravitational time delay effects on cosmic microwave background
  anisotropies,''
  Phys.\ Rev.\  D {\bf 63}, 023504 (2001)
  [arXiv:astro-ph/0008001].
  %%CITATION = PHRVA,D63,023504;%%

%\cite{Seljak:1995ve}
\bibitem{Seljak:1995ve}
  U.~Seljak,
  ``Gravitational lensing effect on cosmic microwave background anisotropies: A
  Power spectrum approach,''
  Astrophys.\ J.\  {\bf 463}, 1 (1996)
  [arXiv:astro-ph/9505109].
  %%CITATION = ASJOA,463,1;%%

%\cite{Hu:2000ee}
\bibitem{Hu:2000ee}
  W.~Hu,
  ``Weak lensing of the CMB: A harmonic approach,''
  Phys.\ Rev.\  D {\bf 62}, 043007 (2000)
  [arXiv:astro-ph/0001303].
  %%CITATION = PHRVA,D62,043007;%%

%\cite{Komatsu:2003fd}
\bibitem{Komatsu:2003fd}
  E.~Komatsu {\it et al.}  [WMAP Collaboration],
  ``First Year Wilkinson Microwave Anisotropy Probe (WMAP) Observations: Tests
  of Gaussianity,''
  Astrophys.\ J.\ Suppl.\  {\bf 148}, 119 (2003)
  [arXiv:astro-ph/0302223].
  %%CITATION = APJSA,148,119;%%

%\cite{Liguori:2005rj}
\bibitem{Liguori:2005rj}
  M.~Liguori, F.~K.~Hansen, E.~Komatsu, S.~Matarrese and A.~Riotto,
  ``Testing Primordial Non-Gaussianity in CMB Anisotropies,''
  Phys.\ Rev.\  D {\bf 73}, 043505 (2006)
  [arXiv:astro-ph/0509098].
  %%CITATION = PHRVA,D73,043505;%%
  
%\cite{Creminelli:2003iq}
\bibitem{Creminelli:2003iq}
  P.~Creminelli,
  ``On non-gaussianities in single-field inflation,''
  JCAP {\bf 0310}, 003 (2003)
  [arXiv:astro-ph/0306122].
  %%CITATION = JCAPA,0310,003;%%

\bibitem{Peebles}
  P.~J.~E.~Peebles,
  ``The Large-Scale Structure of the Universe,''
  (Princeton Series in Physics, 1980) {\it  Princeton Univ Pr 1980, 440p}.
  %%CITATION = NATUA,217,511;%%

%\cite{Rees:1968zz}
\bibitem{Rees:1968zz}
  M.~J.~Rees and D.~W.~Sciama,
  ``Large scale Density Inhomogeneiies in the Universe,''
  Nature {\bf 217}, 511 (1968).

%\cite{Spergel:1999xn}
\bibitem{Spergel:1999xn}
  D.~N.~Spergel and D.~M.~Goldberg,
  ``Microwave background bispectrum. 1. Basic formalism,''
  Phys.\ Rev.\  D {\bf 59}, 103001 (1999)
  [arXiv:astro-ph/9811252].
  %%CITATION = PHRVA,D59,103001;%%

%\cite{Lewis:2006fu}
\bibitem{Lewis:2006fu}
  A.~Lewis and A.~Challinor,
  ``Weak Gravitational Lensing of the CMB,''
  Phys.\ Rept.\  {\bf 429}, 1 (2006)
  [arXiv:astro-ph/0601594].
  %%CITATION = PRPLC,429,1;%%

%\cite{Zaldarriaga:2000ud}
\bibitem{Zaldarriaga:2000ud}
M.~Zaldarriaga,
``Lensing of the CMB: Non-Gaussian aspects,''
Phys.\ Rev.\  D {\bf 62}, 063510 (2000)
[arXiv:astro-ph/9910498].
%%CITATION = PHRVA,D62,063510;%%

%\cite{Creminelli:2004yq}
\bibitem{Creminelli:2004yq}
  P.~Creminelli and M.~Zaldarriaga,
  ``Single field consistency relation for the 3-point function,''
  JCAP {\bf 0410}, 006 (2004)
  [arXiv:astro-ph/0407059].
  %%CITATION = JCAPA,0410,006;%%

%\cite{Cheung:2007sv}
\bibitem{Cheung:2007sv}
  C.~Cheung, A.~L.~Fitzpatrick, J.~Kaplan and L.~Senatore,
  ``On the consistency relation of the 3-point function in single field
  inflation,''
  JCAP {\bf 0802}, 021 (2008)
  [arXiv:0709.0295 [hep-th]].
  %%CITATION = JCAPA,0802,021;%%

  %\cite{Seery:2008ax}
\bibitem{Seery:2008ax}
  D.~Seery, M.~S.~Sloth and F.~Vernizzi,
  ``Inflationary trispectrum from graviton exchange,''
  JCAP {\bf 0903}, 018 (2009)
  [arXiv:0811.3934 [astro-ph]].
  %%CITATION = JCAPA,0903,018;%%

%\cite{Meerburg:2009ys}
\bibitem{Meerburg:2009ys}
  P.~D.~Meerburg, J.~P.~van der Schaar and P.~S.~Corasaniti,
  ``Signatures of Initial State Modifications on Bispectrum Statistics,''
  arXiv:0901.4044 [hep-th].
  %%CITATION = ARXIV:0901.4044;%%

%\cite{Senatore:2009gt}
\bibitem{Senatore:2009gt}
  L.~Senatore, K.~M.~Smith and M.~Zaldarriaga,
  ``Non-Gaussianities in Single Field Inflation and their Optimal Limits from
  the WMAP 5-year Data,''
  arXiv:0905.3746 [astro-ph.CO].
  %%CITATION = ARXIV:0905.3746;%%

%\cite{Hanson:2009kg}
\bibitem{Hanson:2009kg}
  D.~Hanson, K.~M.~Smith, A.~Challinor and M.~Liguori,
  ``CMB lensing and primordial non-Gaussianity,''
  arXiv:0905.4732 [astro-ph.CO].
  %%CITATION = ARXIV:0905.4732;%%


\end{thebibliography}
\end{document}